%% file: main.tex
\documentclass[conference]{IEEEtran}
\makeatletter
\def\mdseries@tt{m}
\makeatother

\usepackage{flushend}

\usepackage[utf8x]{inputenc} 
\usepackage{graphicx}
\usepackage{xcolor}
\usepackage[export]{adjustbox}
\usepackage{mathtools}
\usepackage{multirow}
\usepackage{booktabs}

\usepackage{caption, subfig}

\graphicspath{{images/}}

\include{code-listings}

\usepackage[font=scriptsize,labelfont=normalfont]{caption}

\begin{document}

\title{Going green: optimizing GPUs for energy efficiency through model-steered auto-tuning}

%\author{Ben van Werkhoven}
%\affiliation{
%\institution{Netherlands eScience Center}
%\city{Amsterdam}
%\country{The Netherlands}}
%\email{b.vanwerkhoven@esciencecenter.nl}

% \author{
% \IEEEauthorblockN{Richard Schoonhoven}
% \IEEEauthorblockA{\textit{Centrum Wiskunde \& Informatica}\\
% Amsterdam, the Netherlands \\
% ORCID: }
% \IEEEauthorblockN{Bram Veenboer}
% \IEEEauthorblockA{\textit{Astron}\\
% Amsterdam, the Netherlands \\
% ORCID: 0000-0002-7508-3272}
% \IEEEauthorblockN{Ben van Werkhoven}
% \IEEEauthorblockA{\textit{Netherlands eScience Center}\\
% Amsterdam, the Netherlands \\
% ORCID: 0000-0002-7508-3272}
% \IEEEauthorblockN{Joost Batenburg}
% \IEEEauthorblockA{\textit{Centrum Wiskunde \& Informatica \\ Leiden University }\\
% Amsterdam, the Netherlands \\
% ORCID: 0000-0002-7508-3272}
% }

\author{Richard Schoonhoven\textsuperscript{\rm 1,2}\qquad~Bram Veenboer\textsuperscript{\rm 3}\qquad~Ben~van~Werkhoven\textsuperscript{\rm 1,4}\qquad~K.~Joost~Batenburg\textsuperscript{\rm 1,2}\\
		{\textsuperscript{\rm 1}Computational Imaging Group, Centrum Wiskunde \& Informatica, Amsterdam, Netherlands}\\
		{\textsuperscript{\rm 2}Leiden Institute of Advanced Computer Science, Leiden, Netherlands}\\
		{\textsuperscript{\rm 3}Netherlands Institute for Radio Astronomy (ASTRON), Dwingelo, Netherlands}\\
		{\textsuperscript{\rm 4}Netherlands eScience Center, Amsterdam, Netherlands}\\
		\small{\texttt{\{richard.schoonhoven, k.j.batenburg\}@cwi.nl, veenboer@astron.nl,}}\\ \small{\texttt{b.vanwerkhoven@esciencecenter.nl}}
}

\maketitle
%\IEEEpeerreviewmaketitle

\thispagestyle{plain}
\pagestyle{plain}

\begin{abstract}
Graphics Processing Units (GPUs) have revolutionized the computing landscape over the past decade. However, the growing energy demands of data centres and computing facilities equipped with GPUs come with significant capital and environmental costs. The energy consumption of GPU applications greatly depend on how well they are optimized. Auto-tuning is an effective and commonly applied technique of finding the optimal combination of algorithm, application, and hardware parameters to optimize performance of a GPU application. In this paper, we introduce new energy monitoring and optimization capabilities in Kernel Tuner, a generic auto-tuning tool for GPU applications.
These capabilities enable us to investigate the difference between tuning for execution time and various approaches to improve energy efficiency, and investigate the differences in tuning difficulty.
Additionally, our model for GPU power consumption greatly reduces the large tuning search space by providing clock frequencies for which a GPU is likely most energy efficient.

\end{abstract}

\section{Introduction}

Huge amounts of compute power are powering today’s industrial and scientific applications, at huge energy and environmental costs. Energy is among the largest expenses of supercomputers and data centres, and this consumption will double every four years~\cite{dr_rado_danilak_why_nodate}. The computational demands in deep learning (artificial intelligence) applications have been increasing at a exponential rate, $300{,}000\times$ from 2012 to 2018~\cite{schwartz_green_2019}. The carbon footprint of these applications is a great concern for the environment, as training a single large model produces as much carbon dioxide as five cars in their lifetime, including fuel~\cite{strubell_energy_2019}. In addition, many applications have stringent energy constraints; embedded and automotive systems have limited battery capacity, offshore applications where a connection to the power grid is not possible, and also large-scale scientific instruments, such as the Square Kilometre Array (SKA) built partially in the desert~\cite{dewdney_ska1_2013}. Graphics Processing Units (GPUs) are powering nearly all large-scale AI and HPC applications, and are in large part responsible for the total power consumption of these systems~\cite{pavan_improving_2018, xizhou_feng_power_2005}. For instance, 8.3 MW out of the total 13 MW by the Summit Supercomputer is consumed by its GPUs~\cite{stachowski_autotuning_2020}. There is a clear urgency to improving the energy efficiency of these applications.

While GPUs are relatively energy-efficient processors, energy consumption greatly depends on how well the application is optimized to efficiently use the underlying hardware~\cite{dong_step_2014,li_note_2009}. The optimization of GPU applications is a complex problem that requires finding the best performing combination of many implementation choices and code optimization parameters in a large and discontinuous search space~\cite{ryoo_program_2008, nugteren_cltune_2015, spafford_maestro_2010, lim_autotuning_2017}. As such, auto-tuning, the process of automatically searching for the best performing configuration, is often used to optimize the compute performance of these applications~\cite{grewe_automatically_2011, tomov_dense_2010, zhang_auto-generation_2012, mametjanov_autotuning_2012}.

This has led to the rise of generic GPU code auto-tuners, such as CLTune~\cite{nugteren_cltune_2015}, Kernel Tuner~\cite{vanwerkhoven2019kernel}, Kernel Tuning Toolkit (KTT)~\cite{filipovivc2017autotuning}, and Auto-Tuning Framework (ATF)~\cite{raschTACO}, which facilitate the creation of auto-tuned GPU applications, and support different optimization strategies to accelerate the search process. These frameworks focus on auto-tuning user-defined code parameterizations, which is more generic and powerful than compiler-based auto-tuning~\cite{ashouri2019survey}, because it allows users to tune for entirely different ways to parallelize a computation, with different algorithms to compare, and different data layouts, loop permutations, and code optimizations.
However, none of these generic GPU auto-tuners has built-in support for energy optimization, and the differences between auto-tuning for compute performance and energy efficiency have not yet been studied in detail.

In this paper, we introduce new energy monitoring capabilities in Kernel Tuner, which allows us to use the existing frameworks to study and optimize energy efficiency. We use these capabilities to investigate how different compute performance tuning (lowest kernel runtime) is from energy tuning, and whether the tuning difficulty differs from the perspective of blind optimization algorithms. In addition, we compare two methods for tuning energy efficiency of GPUs; power capping and fixing clock frequencies. Lastly, we introduce a method to efficiently model GPU power consumption, which allows us to significantly narrow the range of clock frequencies to search for the most energy efficient configuration.
All together, we provide a method and open-source tool for tuning GPU applications for both performance and/or energy efficiency. Moreover, these tools can be used for further auto-tuning and high performance computing research.
%contributions

%- generieke tool for GPU energy tuning, maar zeker ook voor het doen van onderzoek naar deze technologie

%antwoorden op de volgende onderzoeksvragen:
%- how different are performance or energy tuning really?
%- how large is the effect of using an external sensor over an internal power sensor? Would the tuner have come to different conclusions?
%- does the effectiveness of search algorithms change for the different objectives?

\section{Related Work}

%write something positive about how beautiful it is that there are so many tools in active development

OpenTuner~\cite{ansel_opentuner_2014} was one of the first generic software auto-tuning frameworks, supporting a number of different search optimization algorithms, but lacks support for tuning individual GPU kernels.
CLTune~\cite{nugteren_cltune_2015} was one of the first of a new breed of generic auto-tuning tools with specific support for tuning GPU kernels written in OpenCL. 
Kernel Tuning Toolkit (KTT)~\cite{filipovivc2017autotuning} is developed specifically to support online auto-tuning and pipeline tuning, which allows for exploration of combinations of tunable parameters over multiple kernels. An interesting feature of KTT is its support for keeping track of hardware performance counters during benchmarking, which can also be used in advanced search strategies~\cite{filipovivc2021using}.
Auto-Tuning Framework (ATF)~\cite{raschTACO} implements a way to generate search spaces, using a chain-of-tree search space structure for efficient storage and fast exploration of constrained search spaces.
HyperMapper~\cite{nardi2019hypermapper} is a tuning framework that focuses on multi-objective optimization and exploitation of user prior knowledge.
Kernel Tuner~\cite{vanwerkhoven2019kernel} is specifically designed to be an easy-to-use and easy to extend tool for the development of tunable GPU kernels, and in particular supports a large selection of search optimization strategies. 
In this paper, we extend Kernel Tuner~\cite{vanwerkhoven2019kernel} with functionality for auto-tuning energy efficiency, which cannot be found in any of the existing generic auto-tuning frameworks.

%quick overview of related work in GPU energy tuning

Research in auto-tuning GPU applications for energy efficiency is still in its infancy, despite spanning more than 12 years of research.
There is no state-of-the-art method for GPU energy tuning, as comparisons between studies or even to a shared baseline are non-existent. The majority of studies only tune individual parameters, e.g. thread block dimensions~\cite{wang_analysis_2010, timm_design_2012, park_performancepower_2013, connors_modeling_2015, lin_auto-tuning_2016, holm_gpu_2020}, or clock frequencies~\cite{mei_measurement_2013, ge_effects_2013, price_optimizing_2016, akiki_energy-aware_2018, fan_accurate_2020, calore_energy-performance_2015}. Only two studies actually combine auto-tuning code optimizations with execution parameters, such as clock frequencies, but only for a single application on a single GPU~\cite{miyazaki_bayesian_2018, coplin_effects_2015}. 
%There is also a severe lack in diversity of the studied hardware as only four studies include non-NVIDIA GPUs~\cite{garzon_approach_2017, grasso_energy_2014, latifi_oskouei_cnndroid_2016, jia_gpu_2015}.

%modeling versus measurement, internal versus external sensor

All generic auto-tuning frameworks use empirical performance measurements, most likely because it is difficult to create generalized performance models that capture the complex system that arises from the combination of hardware and software~\cite{saxe_power-efficient_2010, cameron_energy_2013, procaccianti_green_2015}. Some GPU energy tuning studies use highly-inaccurate performance models, with up to 50\% error, to estimate energy consumption without evaluating the impact of these inaccuracies on the auto-tuning results~\cite{lin_auto-tuning_2016, jia_gpu_2015}. Therefore, most studies take an empirical approach, in particular using the GPU’s internal power sensor~\cite{price_optimizing_2016, akiki_energy-aware_2018, fan_accurate_2020, guerreiro_multi-kernel_2015, li_meterpu_2015, hayes_orion_2016, schiffmann_optimizing_2017}, but also through external power sensors~\cite{suda_mathematical_2013, ren_global_2012, datta_stencil_2008, huang_energy_2009, kamil_auto-tuning_2010, HDEEM} often based on custom-built measurement equipment. Internal power sensors are included in most modern GPUs and can be read by software, e.g., using the NVIDIA Management Library (NVML) for NVIDIA GPUs. Such power sensors are therefore highly accessible, but may suffer from low sampling frequencies and low accuracy~\cite{romein_powersensor_2018}. Some researchers try to compensate for these limitations by measuring individual functions for long
periods of time~\cite{anzt_experiences_2015, price_optimizing_2016, pavan_improving_2018}. This approach, however, is impractical for use in auto-tuners, which often have to benchmark many configurations to find the optimum~\cite{sclocco_auto-tuning_2014}.
As such, Kernel Tuner supports an external power sensor, namely PowerSensor2~\cite{romein_powersensor_2018}, which is accurate within 1\% error and at a sampling frequency of 2.87 kHz. This means that PowerSensor2 is capable of accurately measuring the energy consumption of a kernel without the need to prolong the kernel execution time. We have used PowerSensor2 to validate the power measurements taken using NVML.

%how different is energy tuning really from performance tuning?

Many studies claim that there is a clear difference between the optimization objectives of compute performance and energy efficiency, and that the two require different optimization algorithms and parameters~\cite{holm_gpu_2020, fan_accurate_2020, miyazaki_bayesian_2018, coplin_effects_2015, chaparala_autotuning_2015, krzywaniak_performanceenergy_2019}. However, such claims are often not experimentally verified. The relationship between performance and energy efficiency is complicated, and many authors simply optimize energy efficiency by minimizing the kernel execution time, an approach that is sometimes referred to as race-to-idle~\cite{anzt_experiences_2015}. 
In \cite{choi_roofline_2013}, a model for energy is proposed that predicts that energy usage differs from runtime because energy costs for memory operations cannot be hidden while the algorithm is running. Therefore, energy optimality does not depend solely on optimizing FLOPs, but also on balancing energy usage between memory and compute operations.
In this paper, we aim to experimentally verify the differences between tuning for compute performance and energy efficiency.
%In this paper, we experimentally verify the differences between compute performance and energy efficiency as optimization objectives and evaluate the differences in the optimization difficulty of the resulting search spaces and the effectiveness of search algorithms for either objective.

\section{Methodology}

\subsection{GPU power consumption model}
\label{sec:powermodel}
The energy consumed by a GPU over a time interval $[t_0, t_1]$ is related to its power usage $P(t)$ according to
\begin{equation*}
    E=\int^{t_1}_{t_0}P(t)\;dt.
\end{equation*}
The power consumption $P(t)=V(t)I(t)$ can be determined by measuring the current $I$, and voltage $V$. In practice, one can either approximate the integral numerically by, e.g., trapezoidal integration using the power readings, or simply multiplying the average power consumption by the elapsed time $E=\langle P\rangle(t_1-t_0)$. We employ the latter method in this work, where we take the median power reading for $\langle P\rangle$.

The power consumption of a GPU is affected by several factors, including the workload and operating frequency of the GPU. The workload is implementation dependent, and in most cases can be optimized by tuning kernel parameters, or by changing the kernel code. Furthermore, different GPU models contain different components, such as memory and chips, that operate at certain clock frequencies which can vary at runtime. These operating frequencies are commonly taken as is.

Throughout this work, we use a variety of GPUs with distinct architectures. Moreover, even within one architecture (e.g. the Ampere architecture) we cannot assume that the energy characteristics of two different models are identical. The Tesla A100 and RTX A4000 GPUs for instance use a different chip (GA100 versus GA102), are produced at a different process size (7 nm versus 8 nm), and have a very different mix and number of execution units. Moreover, the Tesla A100 has HBM2e memory, while the RTX A4000 uses GDDR6. The NVIDIA drivers currently do not expose an option to tune the clock frequency of the HBM memory. For the RTX A4000 and a compute-bound kernel, we measured only a marginally lower energy consumption when reducing the memory clock frequency. Therefore, we consider solely the graphics clock (core) frequency in this work.

Contemporary GPUs usually operate at a base core frequency and can boost up to a certain turbo frequency to increase performance, but only when the temperature and power consumption of the device allows for it. This technique is commonly referred to as Dynamic Voltage Frequency Scaling (DVFS). Price et al.~\cite{price_optimizing_2016} showed a relation between core frequency and the voltage required to operate on a given frequency, and a power consumption model is given by
\begin{equation}
    P_{gpu} = P_{static} + N_c C f V^2,
    \label{eq:powersum}
\end{equation}
where $C$ is load capacitance, $N_c$ the number of switches, $f$ is frequency, and $V$ is voltage. $V$ typically increases with $f$. Consequently, the turbo frequency may be good for performance, but not necessarily for energy efficiency.

To steer frequency tuning, we fit a GPU power consumption model to data in section \ref{sec:powermodelling}, using a non-linear least squares approach (Levenberg-Marquardt algorithm~\cite{more1978levenbergmarquardt}).

\subsection{Energy measurements in Kernel Tuner}

\begin{figure}[t]
    \centering
    \includegraphics[width=0.9\columnwidth]{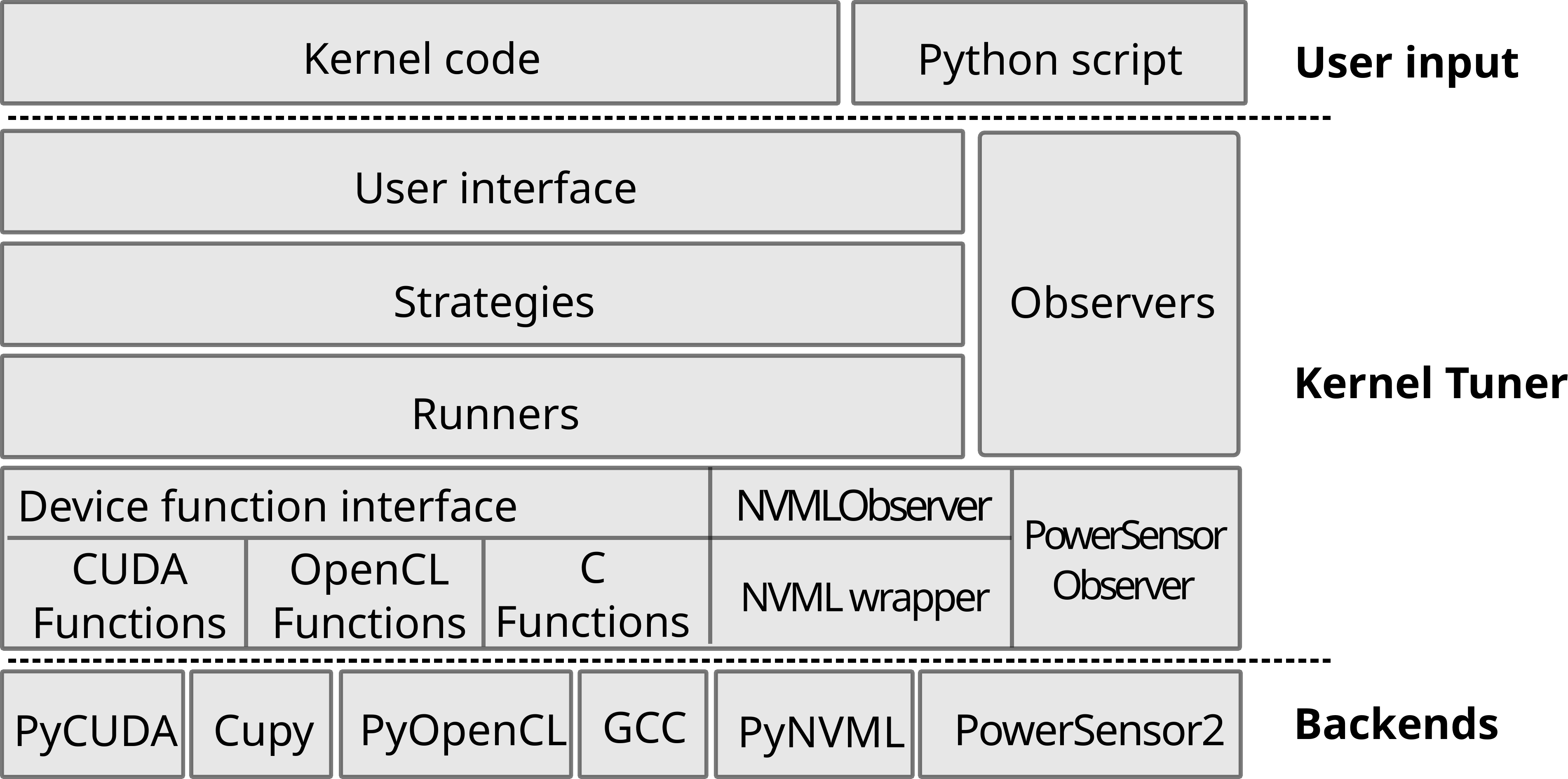}
    \caption{Extended software architecture of Kernel Tuner.}
    \label{fig:architecture}
    \vspace{-0.6em}
\end{figure}

%Expand layered image with stacks at the bottom or next to it for NVML, and PowerSensor
We introduce several new features in Kernel Tuner to acquire energy measurements of GPU kernel executions, namely observers, user-defined metrics, and custom tuning objectives. The software architecture and basic functionality of Kernel Tuner is described in \cite{vanwerkhoven2019kernel}, and a diagram of software hierarchy can be found in Figure \ref{fig:architecture}. An observer can be implemented to execute functions and can extend results obtained during benchmarking before, during and after kernel execution. For the experiments in this work, we implemented the \verb=NVMLObserver= and \verb=PowerSensorObserver= in Kernel Tuner.

%Kernel Tuner uses a layered software architecture and the main design philosophy is that it should be easy to use and easy to extend. The software architecture and basic functionality of Kernel Tuner is described in \cite{vanwerkhoven2019kernel}.

\subsubsection{PowerSensorObserver}

To facilitate accurate energy measurements at high sampling frequency, we implemented the \verb=PowerSensorObserver= (using PyBind11\footnote{https://pybind11.readthedocs.io/en/stable/}) as an interface to PowerSensor2~\cite{romein_powersensor_2018}.
The user can select this observer to record power and/or energy consumption of kernel configurations during auto-tuning. This allows Kernel Tuner to accurately determine the power and energy consumption of all kernel configurations it benchmarks during auto-tuning.

% PowerSensor2~\cite{romein_powersensor_2018} is a custom-built power measurement device for PCIe devices that 
% intercepts the device power with current sensors and transmits the data to the host over a USB connection. The main advantage of using PowerSensor2 over the GPU's built-in power sensor is that PowerSensor2 reports instantaneous power consumption with a very high frequency (about 2.8 KHz). PowerSensor2 comes with an easy-to-use software library that supports various forms of power measurement. We have created a simple interface using PyBind11\footnote{https://pybind11.readthedocs.io/en/stable/} to the PowerSensor library to make it possible to use it from Python.

% Kernel Tuner implements a PowerSensorObserver specifically for use with PowerSensor2, that can be selected by the user to record power and/or energy consumption of kernel configurations during auto-tuning. This allows Kernel Tuner to accurately determine the power and energy consumption of all kernel configurations it benchmarks during auto-tuning.

%\medskip

\subsubsection{NVMLObserver}

Measurements with the PowerSensor2 require wiring external hardware to a GPU, and the sensor is not available to most users, the bulk of our measurements will be performed using NVIDIA's internal sensors.
The NVIDIA Management Library (NVML)~\cite{nvml} can be used for power measurements on almost all NVIDIA GPUs, so using this library is much more accessible to end-users compared to solutions that require custom hardware, such as PowerSensor2.
To this end we implemented the \verb=NVMLObserver= in Kernel Tuner, which allows the user to observe the power usage, energy consumption, core and memory frequencies, core voltage and temperature as reported by NVML.
%To facilitate the interaction with NVML, Kernel Tuner implements a thin wrapper that abstracts some of the intricacies of NVML into a more user friendly and Pythonic interface. The NVMLObserver is implemented on top of this interface.

As opposed to PowerSensor2, the power usage reported by NVML has a significantly lower temporal resolution. Furthermore, NVML only reports a time-averaged power consumption rather than instantaneous power consumption~\cite{burtscher_measuring_2014}. 

\begin{figure}
    \centering
    \includegraphics[width=\columnwidth]{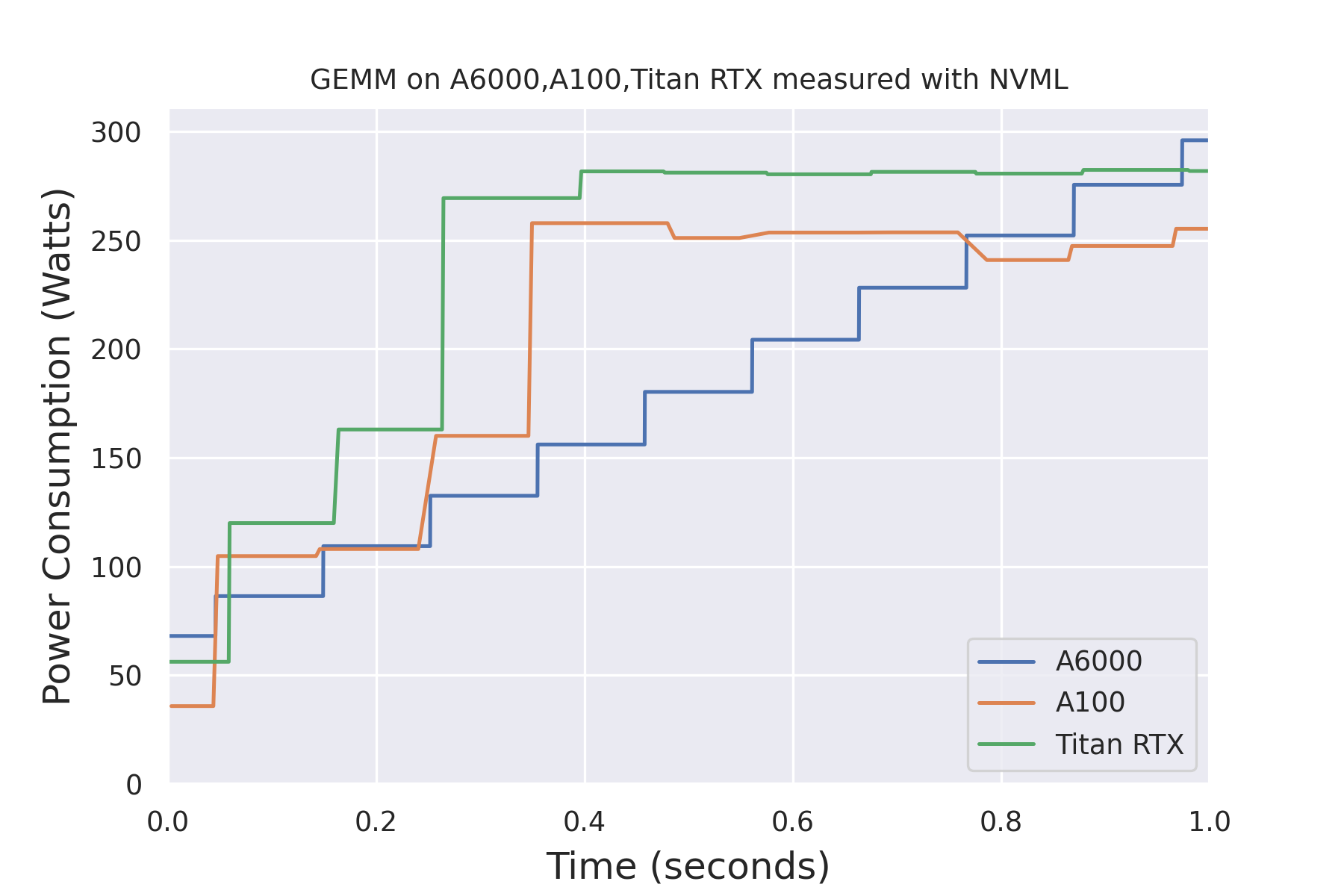}
    \caption{NVML power readings while executing matrix multiplication kernel (GEMM) over time on three different GPUs.}
    \label{fig:nvml_test}
    \vspace{-0.6em}
\end{figure}

Figure~\ref{fig:nvml_test} shows the GPU power consumption over time as reported by NVML, while continuously executing a matrix multiplication kernel (GEMM see section \ref{sec:setup}) for one second.
The jumps in the graph are caused by the fact that the time-averaged value reported by NVML only refreshes at a frequency of about 10~Hz (9.75~Hz on RTX A6000, 14.5~Hz on Tesla A100, and 12.4~Hz on Titan RTX). 
We can see that on the Titan RTX and Tesla A100, the power consumption as report by NVML stabilizes after about 0.3 seconds into the run. For the RTX A6000, power consumption gradually ramps up until hitting the Thermal Design Power (TDP) right before the end of our 1-second interval. 

To ensure that the NVML power measurements in Kernel Tuner more accurately reflect the power consumption of the kernel, the \verb=NVMLObserver= executes the kernel repeatedly for a user-specified duration (1 second by default), and takes the final energy measurement, thereby ensuring a more accurate measurement with NVML.
The downside of this approach is that it significantly increases benchmarking time.

\subsection{Tunable parameters and objectives for energy tuning}

Using application-specific clock frequencies is one of the most common approaches to tuning energy efficiency on GPU systems. Recently, Krzywaniak and Czarnul~\cite{krzywaniak_performanceenergy_2019} have shown promising results with setting application-specific power limits, also called {\em power capping}, to optimize energy consumption. For this work, we have implemented support in Kernel Tuner for users to tune their applications under different clock frequencies and power limits.
Specifically, NVML tunable parameters, such as \verb|nvml_gr_clock|, \verb|nvml_mem_clock|, and \verb|nvml_pwr_limit|, can be set using Kernel Tuner.
Note that changing these settings requires root privileges on most systems. 
As such, these features may not be available to all users on all systems.

%\subsection{User-defined metrics}

Lastly, to perform energy tuning, we need to specify metrics that we aim to minimize or maximize. Using the aforementioned observers, we can collect power readings (in Watts) during kernel execution.
Furthermore, Kernel Tuner's flexible \emph{user-defined metrics} allows us to define other metrics such as \emph{compute performance} in floating point operations per second (GFLOP/s).
This allows us to define \emph{energy efficiency} as GFLOPs/W (same as GFLOP/J) which is a measure of the energy used to perform a billion floating point operations.

\begin{table}[]
    \scriptsize
    \centering
    \begin{tabular}{l|l|r|r|r|r}
    \toprule
    GPU & Architecture & Cores & Bandwidth & Peak SP & TDP (W) \\
    \midrule
    RTX A4000 & Ampere (GA104) & 6{,}144 & 448 & $19{,}170$ & 140 \\
    RTX A6000 & Ampere (GA104) & 10{,}752 & 768 & $38{,}709$ & 300 \\
    Tesla A100 & Ampere (GA100) & 6{,}912 & 1{,}555 & $19{,}500$ & 250 \\
    Tesla V100 & Volta (GV100) & 5{,}120 & 900 & $14{,}028$ & 250 \\
    Titan RTX & Turing (TU102) & 4{,}608 & 672 & $16{,}312$ & 320 \\
    \bottomrule
    \end{tabular}
    \caption{GPUs used in our experiments. Bandwidth in GB/s. Peak compute performance in GFLOP/s. TDP in Watts.}
    \label{tab:gpu-properties}
\end{table}

\section{Experimental setup}
\label{sec:setup}
To investigate energy tuning on GPUs, we run several real-world applicable kernel programs, on a few different GPUs available in either the DAS-6 cluster (Turing and Ampere architecture)~\cite{bal2016das}, or in the LOFAR COBALT-2 correlator system (Tesla V100)~\cite{broekema2018cobaltcorrelator}. Table~\ref{tab:gpu-properties} lists the properties of these GPUs.
In addition to the widely-used GEMM kernel, we validate our results on several computationally expensive radio astronomy kernels currently processing data for the Low Frequency Array (LOFAR) radio-telescope~\cite{vanhaarlem2013_lofar}. These kernels will be used in section \ref{sec:practicalgain} to determine the practically obtained energy reduction for a real-world application. All kernels are compute-bound, except for the TDD kernel which is memory-bound. For the experiments in this section,

{\bf GEMM} (Generalized dense matrix–matrix multiplication) is one of the most widely-used kernels across many application domains, including neural networks. Here we perform the calculation $C=\alpha A\cdot B+\beta C$ for $4096\times 4096$ matrices $A,B,C$, and constants $\alpha$ and $\beta$. We use the highly-tunable OpenCL implementation available in CLBlast~\cite{nugteren_clblast_2018}.

The CLBlast GEMM kernel can be tuned with many parameters, here we summarize the most important ones:
\begin{itemize}
\item $M\sb{wg}$, $N\sb{wg}$, and $K\sb{wg}$ represent the total size of the tile processed by a single thread block in the M, N, and K matrix dimensions.
\item $M\sb{dimC}$ and $N\sb{dimC}$ are the thread block dimensions in M and N.
\item $SA$ and $SB$ can be used to enable or disable using shared memory as a software managed cache for matrix A and matrix B. 
\item $M\sb{vec}$ and $N\sb{vec}$ are the vector widths for loading and storing to global memory, $M\sb{vec}$ is used for matrices A and C, and $N\sb{vec}$ for matrix B.
\item $K\sb{wi}$ is the unrolling factor used for the loop over K.
\end{itemize}
While the GEMM kernel can use several code optimizations, none of the code optimizations have been introduced to optimize the kernel specifically for energy efficiency.
All tunable parameters combined describe a large space, of which many portions are restricted. Using the parameters employed by CLBlast, the search space consists of 17472 valid kernel configurations, that will all be compiled and benchmarked when performing an exhaustive search. However, when we add additional tunable parameters for energy tuning, such as a power limit or clock frequency, the search space grows combinatorially from a grid search perspective. For example, if we want to tune all parameters in the search space in combination with 7 different clock frequencies, the total size of the search space becomes $17{,}472\times7=122{,}304$.

{\bf LOFAR Correlator} is the correlator application used for real-time processing of LOFAR (Low Frequency Array) data~\cite{vanhaarlem2013_lofar}. It combines measurements from the radio telescope into a data product to be processed further by other (offline) processing pipelines (see other kernels).
%For time-domain astronomy, such a pipeline typically contains a \emph{dedispersion} algorithm (see below for details). For an astronomical imaging pipeline, \emph{gridding} and \emph{degridding} are typically the most time-consuming sub parts (see below).
%The LOFAR Correlator software is currently used in production on the COBALT 2.0 GPU cluster~\cite{broekema2018cobaltcorrelator}. This cluster is based on Tesla V100 GPUs, while the software originates from the original COBALT cluster, which was based on Tesla K40c GPUs.
The correlator kernel was tuned by hand for the Kepler architecture, e.g. by unrolling loops and using fixed block and grid dimensions. Consequently, there is only a single tuning parameter left: \verb=NR_STATIONS_PER_THREAD=. This parameter is used to choose between one of four different kernels.

\begin{figure}
    \centering
    \newcommand{\wdth}{0.5}
    \captionsetup[subfigure]{labelformat=empty}
    \includegraphics[width=\wdth\textwidth, right]{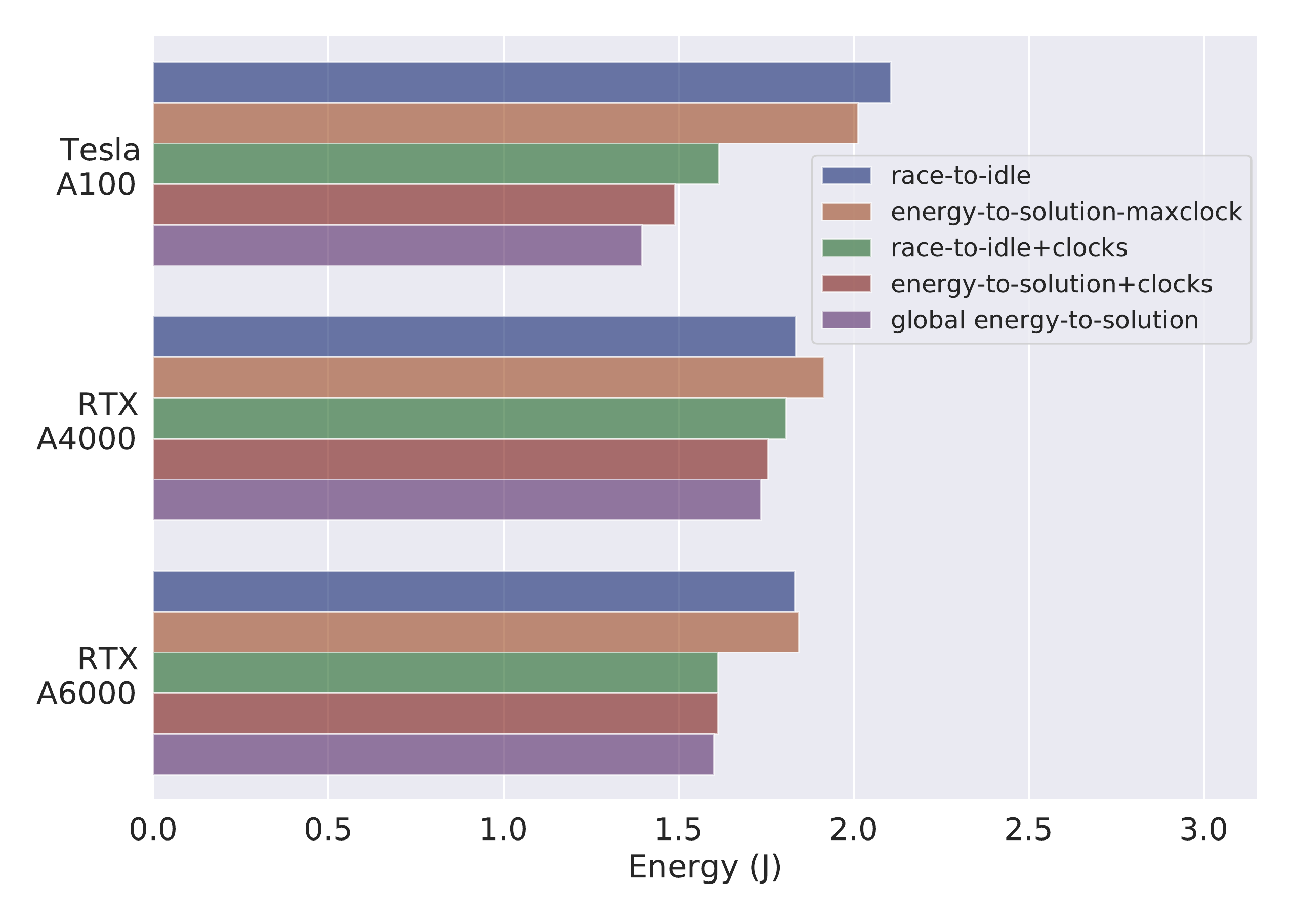}\\
    \vspace{-0.8em}
    \includegraphics[width=0.495\textwidth, right]{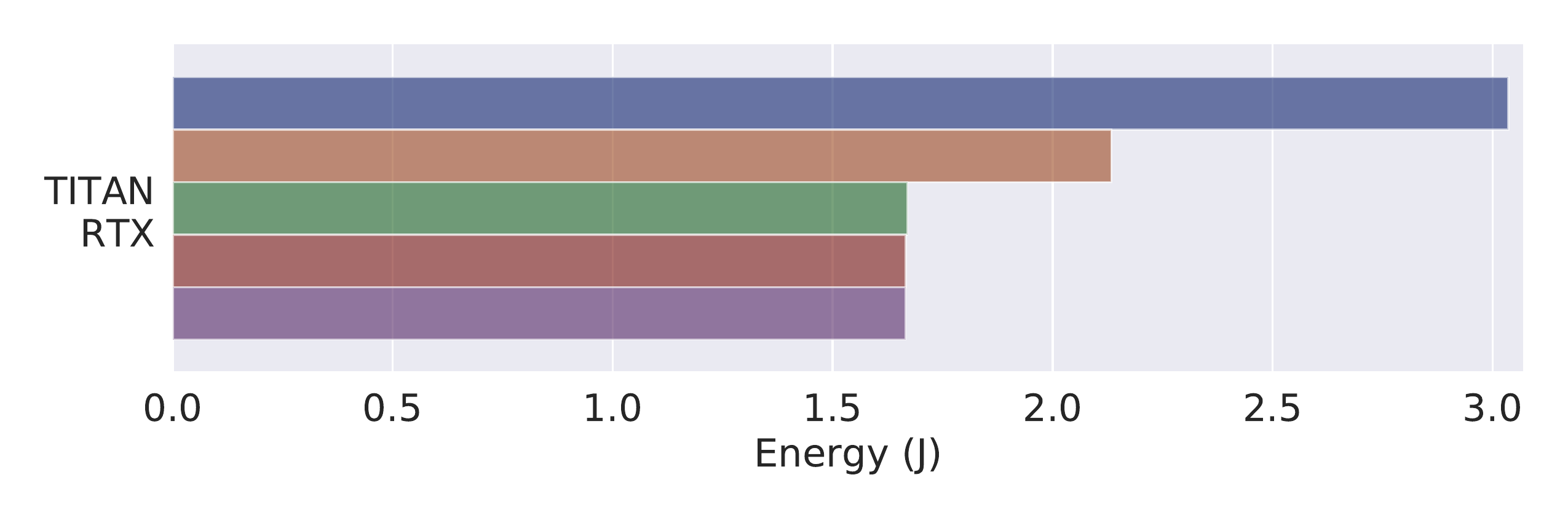}
    \caption{
    \textbf{GEMM:} Lowest energy configuration for the Tesla A100, RTX A4000, RTX A6000, and TITAN RTX GPUs for the  \emph{race-to-idle}, \emph{energy-to-solution-maxclock}, \emph{race-to-idle+clocks}, \emph{energy-to-solution+clocks}, and \emph{global energy-to-solution} tuning methods. The energy measurements for the TITAN RTX were acquired using PowerSensor2, the others using NVML.}
    \label{fig:first_optim_time_freq_max}
    \vspace{-0.6em}
\end{figure}

\begin{figure*}
    \centering
    \newcommand{\hgth}{0.392}%0.372 is max to get more lines
    \newcommand{\bspc}{0.0}
    \newcommand{\hspc}{-0.0}
    \centering
    \captionsetup[subfigure]{labelformat=empty}
    \includegraphics[height=\hgth\textwidth]{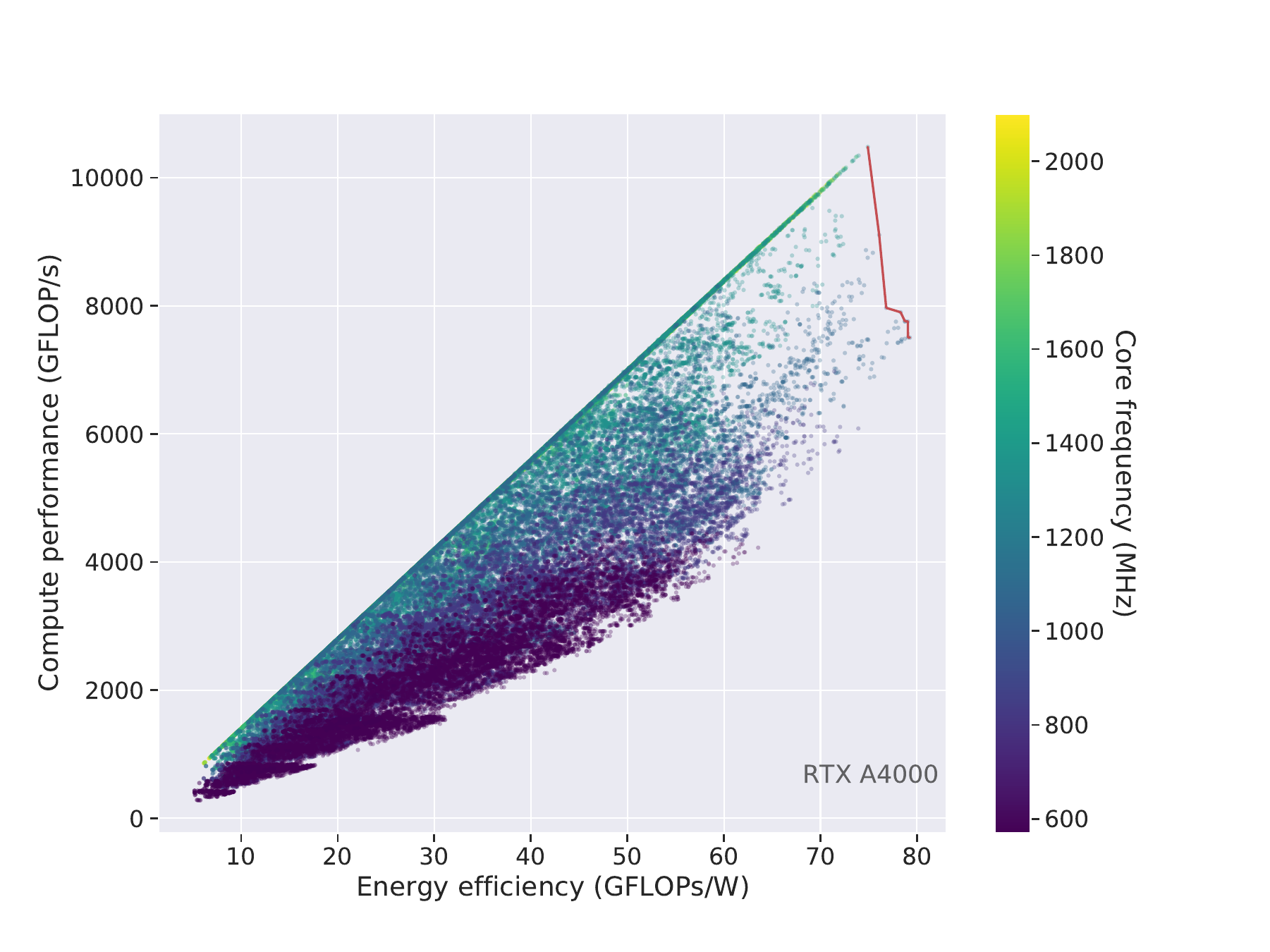}
    \hspace{-3.0em}
    \includegraphics[height=\hgth\textwidth]{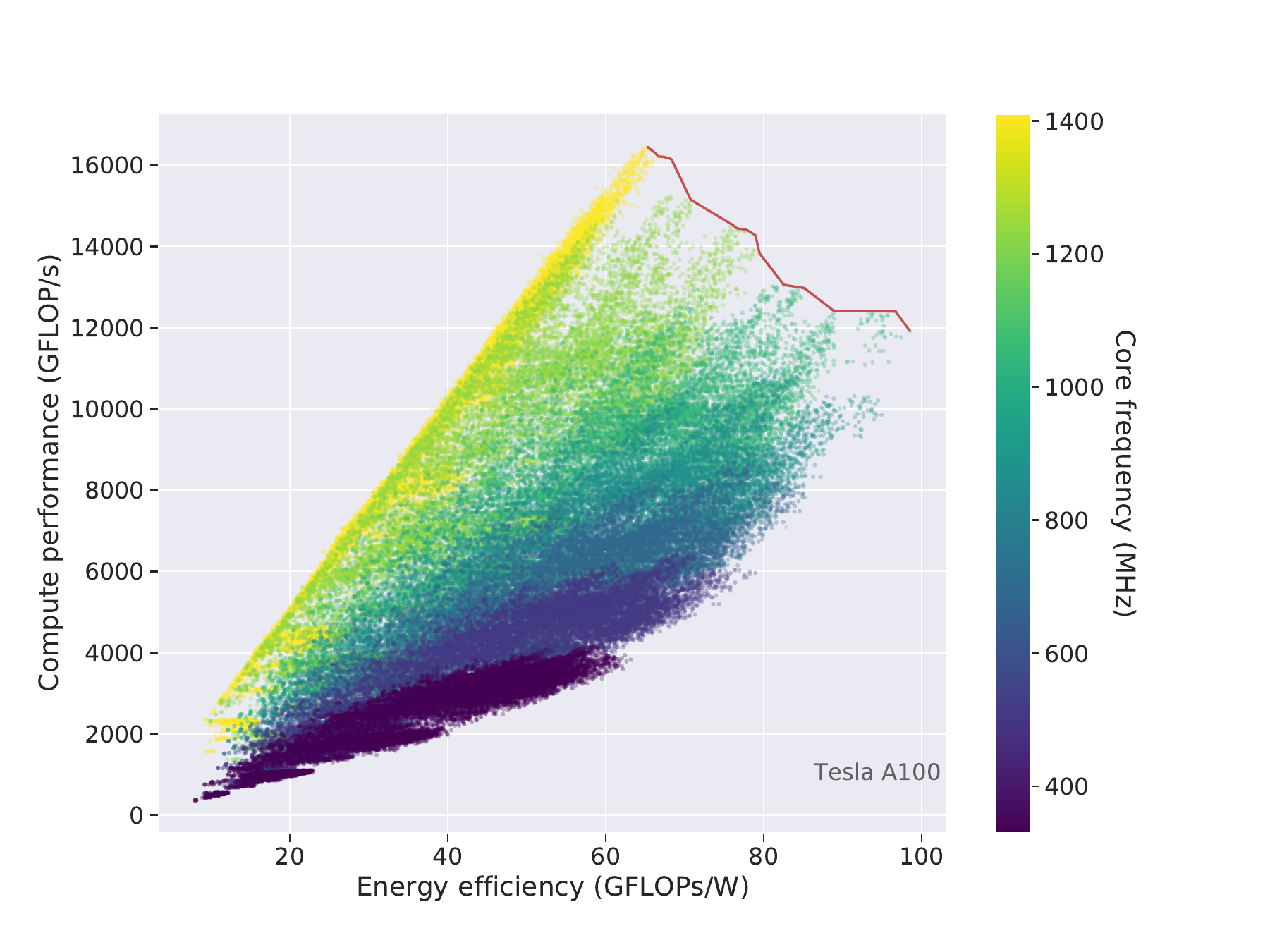}
    \caption{Kernel speed (GFLOP/s) over energy efficiency (GFLOPs/W) for all GEMM configurations for the RTX A4000 (left) and Tesla A100 (right). The red line indicates the Pareto front, i.e., neither performance or efficiency can be improved without decreasing the other. The points are coloured according to the core frequency.}
    \label{fig:pareto_time_over_energy}
    \vspace{-0.6em}
\end{figure*}

{\bf TCC} (Tensor-Core Correlator) is similar to the LOFAR correlator, leveraging the Tensor Cores of contemporary NVIDIA GPUs~\cite{romein2021_tcc}. Tensor Cores are mixed-precision compute units that operate on matrix-like inputs. By using these compute units, the Tensor-Core correlator is both much faster and much more energy-efficient compared to previous correlators. This kernel is hand-tuned and uses fixed thread block dimensions. There is one tuning-parameter: \verb=PORTABLE=, which determines whether the output is written using asynchronous writes (not supported on all GPUs) or via shared memory.% -- the \emph{portable} option.

{\bf IDG} (Image-Domain Gridding) is an algorithm for radio astronomical imaging, of which the \emph{gridder} and \emph{degridder} kernels are the most compute intensive.
%Unlike traditional gridding and degridding,
IDG moves the computation (which resembles convolution) from the \emph{frequency domain} to the \emph{image domain} by introducing \emph{subgrids} and Fourier transformations for processing input data in smaller subsets~{\cite{veenboer2020-ascom, veenboer_radio-astronomical_2019}}.
%Since subgrids are independent of each other, this allows for a highly parallel implementation, which performs excellent on accelerators~{\cite{veenboer2020-ascom, veenboer_radio-astronomical_2019}}.
The GPU implementation of the gridder has the following tuning parameters:
\verb=BLOCK_SIZE_X=, the number of threads in a thread block; \verb=UNROLL_PIXELS=, the number of pixels to process by a thread; \verb=NUM_BLOCKS=, the number of threads blocks per SM; \verb=USE_EXTRAPOLATE=, option to reduce the number of trigonometric operations, at the cost of having to perform more fused multiply-add operations. The degridder kernel has the same options, except for \verb=UNROLL_PIXELS=.

{\bf Dedispersion} is used in time-domain astronomy to detect transient effects (e.g. fast radio bursts) and pulsars. The signal received by the telescope is dispersed (shifted) in time of the frequency band, and dedispersion is needed to correct for this. Dedispersion can either be performed in the \emph{time domain} (TDD), or in the \emph{Fourier domain} (FDD)~\cite{bassa2022_fdd}. TDD has two tuning parameters:  \verb=SAMPS_PER_THREAD=, controls the number of samples to be processed per thread; \verb=USE_TEXTURE_MEM=, whether to use texture memory as a cache when loading input data. FDD has the following tuning parameters: \verb=NFREQ_BATCH_GRID= and \verb=NDM_BATCH_GRID= control the number of input samples to process per kernel invocation; \verb=NCHAN_BATCH_THREAD=, the number of input samples (in the frequency dimension) that every GPU thread processes; \verb=USE_SHARED_MEMORY=, use shared memory as software-managed cache when reading input data; \verb=USE_EXTRAPOLATE=, reduces the number of trigonometric operations (same as for IDG, see above.).

\begin{figure*}
    \centering
    \newcommand{\wdth}{0.34}
    \newcommand{\bspc}{0.0}
    \newcommand{\hspc}{-0.25}
    \captionsetup[subfigure]{labelformat=empty}
    \subfloat{\makebox[\bspc\textwidth][c]{
    \subfloat{\includegraphics[width=\wdth\textwidth]{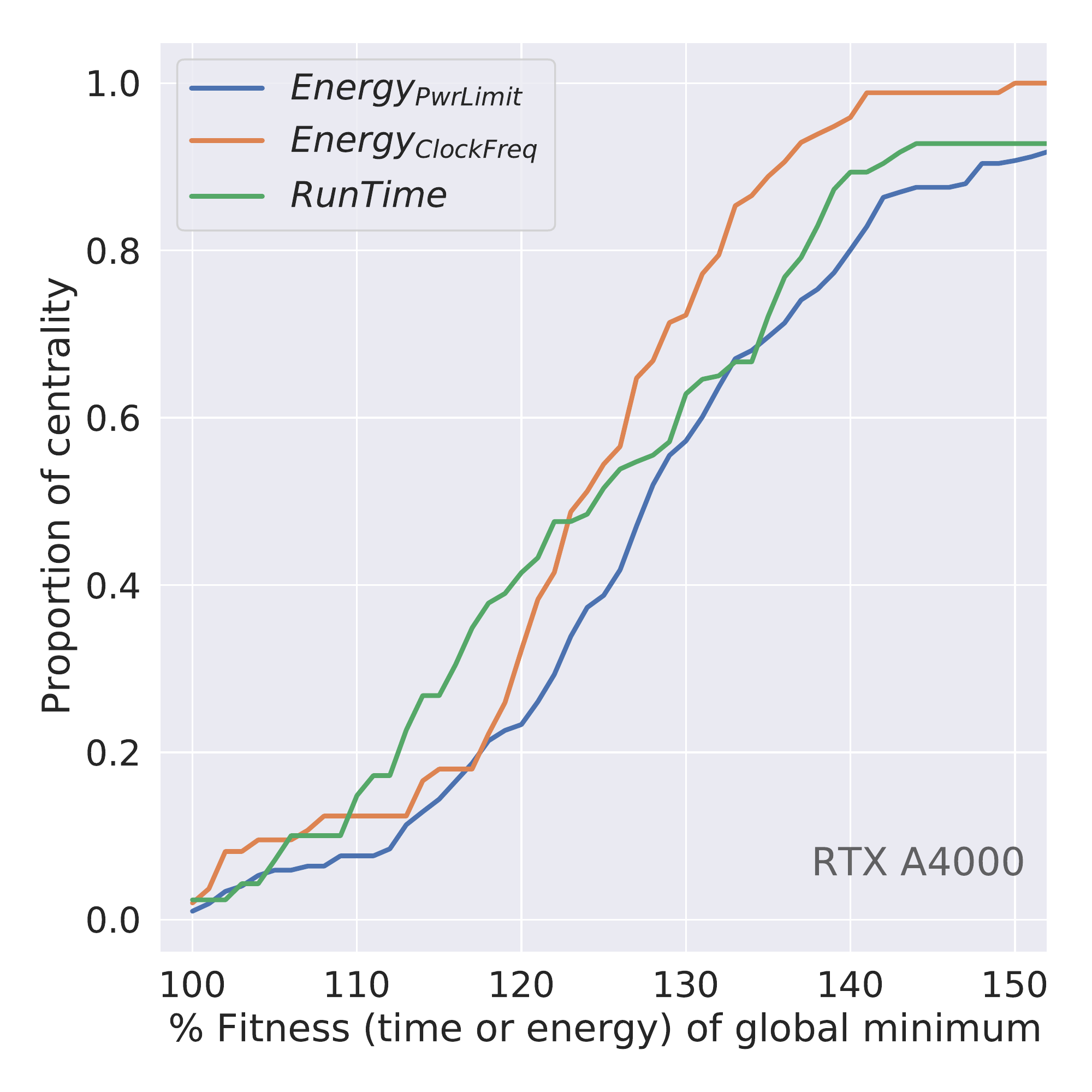}}
    \hspace{\hspc cm}
    \subfloat{\includegraphics[width=\wdth\textwidth]{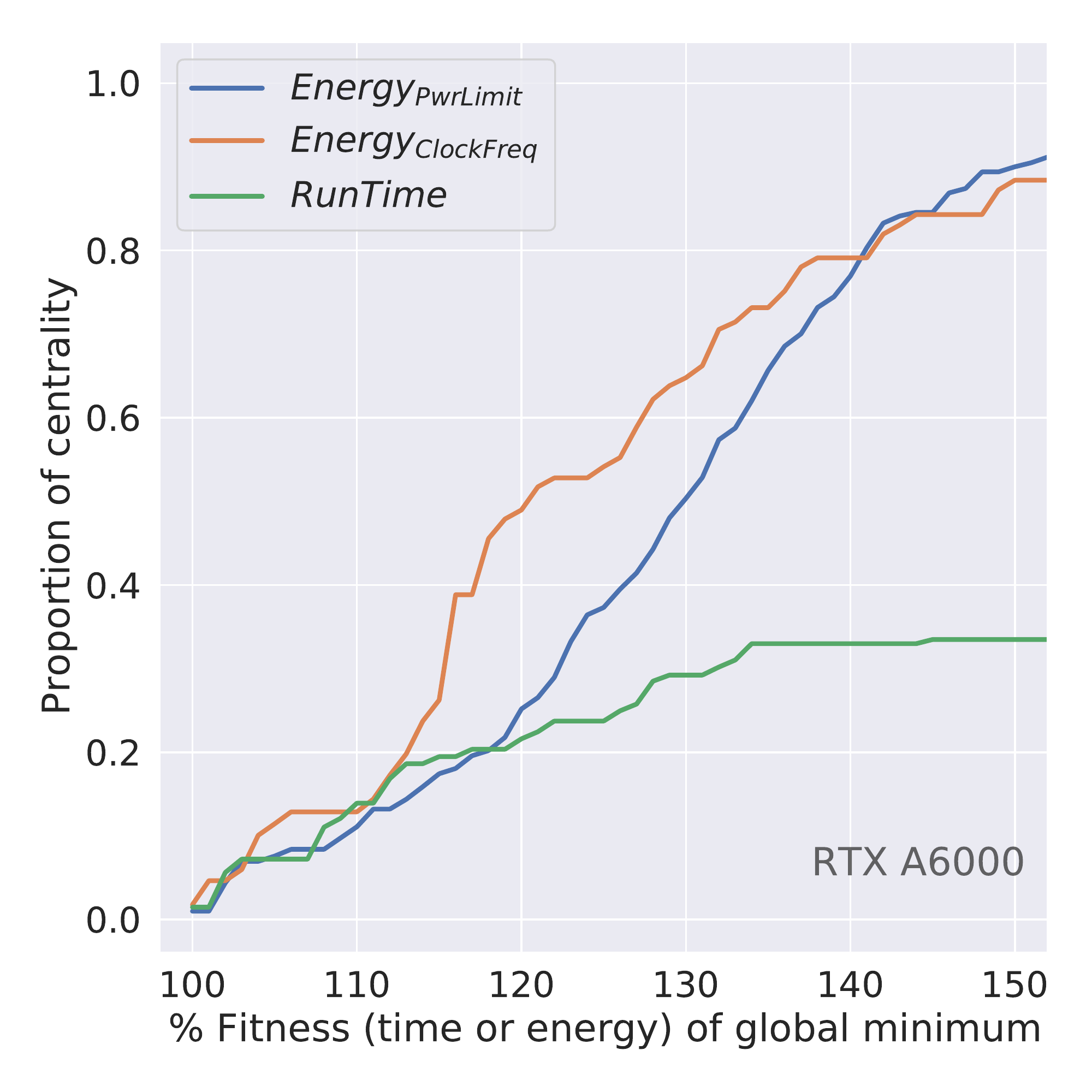}}
    \hspace{\hspc cm}
    \subfloat{\includegraphics[width=\wdth\textwidth]{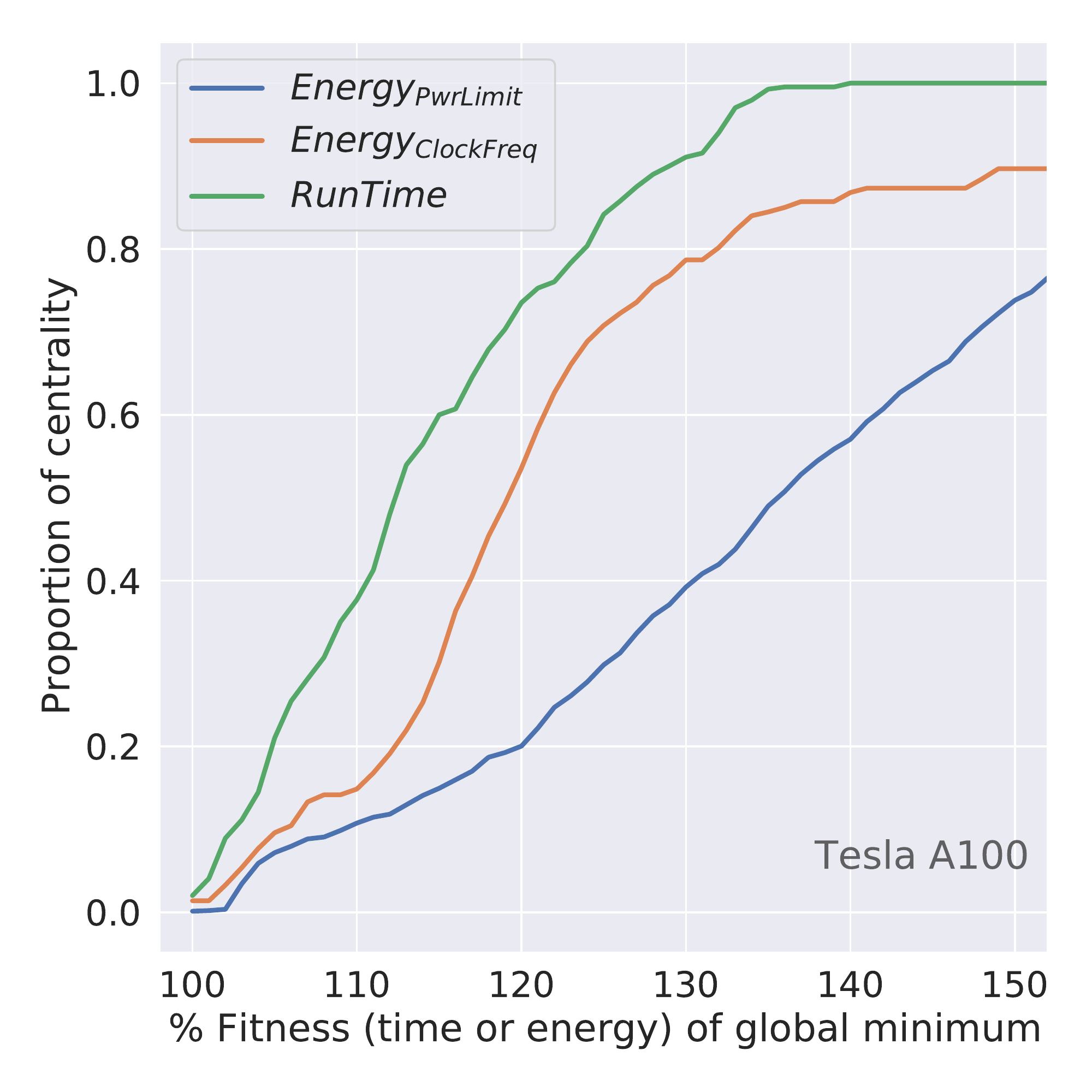}}
    }}
    \caption{Proportion of centrality for tuning execution time, energy tuning (power limit), and energy tuning (clock frequency) for the RTX A4000, RTX A6000, and Tesla A100 GPUs.}
    \label{fig:FFG_GEMM}
    \vspace{-0.6em}
\end{figure*}

\section{Experimental results}

\subsection{Impact of energy tuning versus race-to-idle}
In this section, we experimentally answer whether auto-tuning for energy efficiency (global energy-to-solution) is different from auto-tuning for the lowest kernel runtime across all clock frequencies (race-to-idle).
Furthermore, we report the lowest energy configuration at max clocks.
We compare with a practical compromise where we first tune for time, and then select a clock frequency for the best energy efficiency. We call this last approach {\em race-to-idle+clocks}. Conversely, we also consider \emph{energy-to-solution+clocks} where we fix the frequency at the base clock frequency, tune for energy, and then select a clock frequency to further maximize energy efficiency.

In Figure~\ref{fig:first_optim_time_freq_max}, we show the lowest energy configuration in the GEMM search space with each of the aforementioned methods across several GPUs. For the TITAN RTX we used the PowerSensor2 measurements to validate the findings. We use relatively widely spaced equidistant samples from the range of supported SM clock frequencies (7-points) due to the high cost of obtaining all measurements (9 days per GPU).

First, Figure~\ref{fig:first_optim_time_freq_max} shows that the fastest configuration returned by race-to-idle is not the most energy efficient for any of the GPUs. Second, for most GPUs, the energy usage of the configurations found by race-to-idle+clocks and energy-to-solution+clocks are close to the global lowest energy configuration, but they never have the same parameters. Note that for race-to-idle+clocks, we first tuned for time with the clock frequency fixed to the maximum, before tuning only the clock frequency for energy efficiency.
%When repeating the experiment by tuning for time with the clock frequency fixed to the base clock frequency, the same configurations were found.

The exception is the Tesla A100, where we see a gap in energy usage between all five methods. This means that there is a particular combination of tunable parameter values that results in a configuration that is more energy-efficient than anything returned by the two-step optimization approaches. In other words, to find the global optimum in terms of energy-to-solution it is necessary to search the combined configuration space of all tunable parameters, including clock frequencies.

Our experimental results show that auto-tuning the GEMM kernel for energy efficiency does not lead to the same optimal configuration as tuning for time, as all five methods produce different configurations, with a different energy usage. This raises the question of how kernel speed and energy efficiency are related. In Figure \ref{fig:pareto_time_over_energy} we plot the compute performance in GFLOP/s for every GEMM configuration over energy efficiency in GFLOPs/W, together with the Pareto front in red. By looking at the points on the Pareto front for the RTX A4000 and Tesla A100, we see that the trade-off between speed and energy efficiency differs between GPUs. For the RTX A4000, a speed reduction of $28.4\%$ leads to an increase in energy efficiency of just $5.8\%$. However, for the Tesla A100, a speed reduction of $27.5.\%$ leads to an increase in energy efficiency of $50.9\%$. 
Therefore, the trade-off between kernel runtime and energy usage is GPU specific.

Overall, our results show that, for the GEMM kernel, tuning for lowest energy leads to different configurations than tuning for lowest execution time. However, depending on the GPU, it may be sufficient to treat the optimization as a two-stage optimization problem; first optimizing for minimal energy with a fixed clock frequency, and then optimizing for the most energy efficient frequency, can result in close to optimal energy efficiency on certain GPUs.
%However, for the Tesla A100, it is necessary to treat energy tuning as an optimization problem on the combined configuration space of all tunable parameters.
%It should be noted that, while the GEMM kernel already contains many tunable parameters, none of these parameters have been introduced to optimize the code for energy efficiency in particular. 

\subsection{Speed vs energy: tuning difficulty of optimization spaces}
\label{sec:searchspace}
Tuning a kernel for energy typically requires a larger search space compared to tuning only for execution time. For energy, the search space is typically enlarged with tunable parameters such as clock frequency, or power limit, and possibly other specific optimizations that affect energy usage (e.g. the use of shared memory).
This raises the question whether the search space for energy tuning, compared to tuning execution time, is only larger, or whether energy is actually harder to optimize with optimization algorithms.

The \emph{proportion of PageRank centrality}~\cite{schoonhoven2022} quantifies search difficulty for blind optimization algorithms. Here, a \emph{fitness flow graph} (FFG) is created where all the points in the search space are represented as nodes, and a directed edge from a node to its neighbour is added if the neighbour has better fitness (energy or time). A random walk across the FFG has the property that it mimics a randomized first-improvement local search algorithm. The PageRank centrality of a local minimum in the FFG is the proportion of arrivals in that minimum for a random walk, i.e., the proportion of arrivals of a first-improvement local searcher during optimization. Since local searchers terminate in local minima, the proportion of centrality metric considers the fraction of centrality of ``suitably good'' local minima, among all minima in the space. In other words, it gives the expected fraction of local search terminations in ``good'' local minima. If near-optimal minima have high centrality, a local searcher will find a close to optimal solution in fewer evaluations. Here, ``suitably good'' means that the fitness of the minimum is within $p\cdot f_{optimal}$ for some $p\geq 1$.

\begin{figure*}
    \centering
    \includegraphics[width=\textwidth]{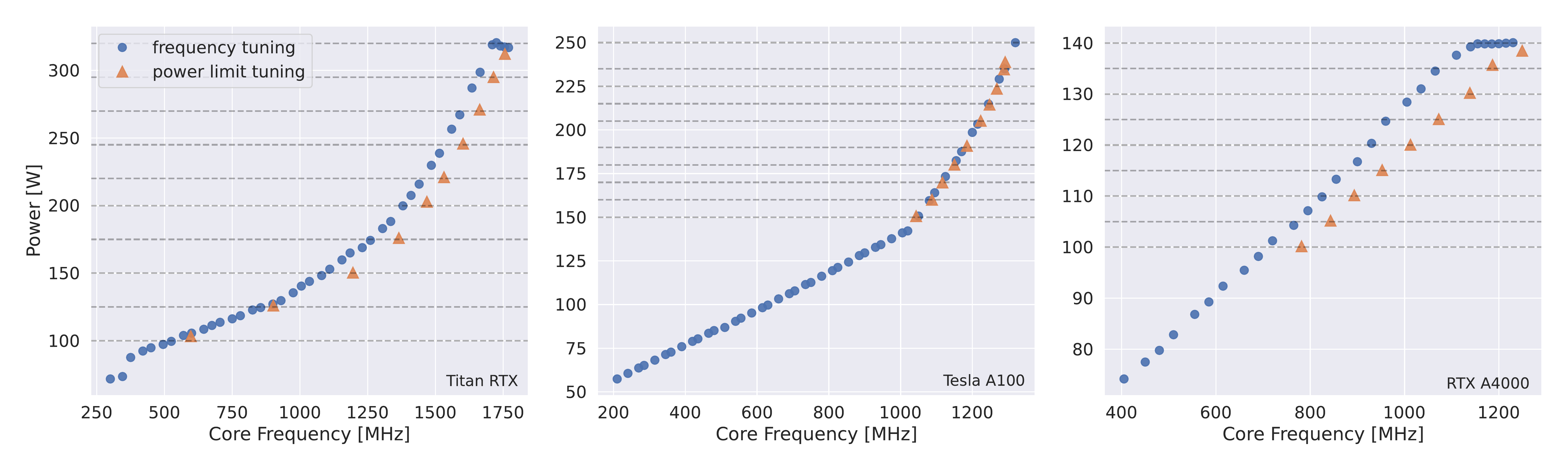}
    \caption{Tuning using a power limit (triangles) versus tuning using frequency (circles) for TITAN RTX (left), Tesla A100 (middle) and RTX A4000 (right) for a synthetic workload that fully occupies the GPU. For all three GPUs, the power consumption coincides with the configured power limit (indicated with the dashed lines).  Moreover, we observe that for this workload, the TITAN RTX and RTX A4000 can not sustain their maximum advertised turbo clock frequency of 1770 MHz and 1560 MHz, respectively.
    }
    \label{fig:power-vs-frequency-tuning}
    \vspace{-0.3em}
\end{figure*}

\begin{figure}
    %\centering
    \newcommand{\wdth}{0.45}
    \includegraphics[width=\wdth\textwidth, center]{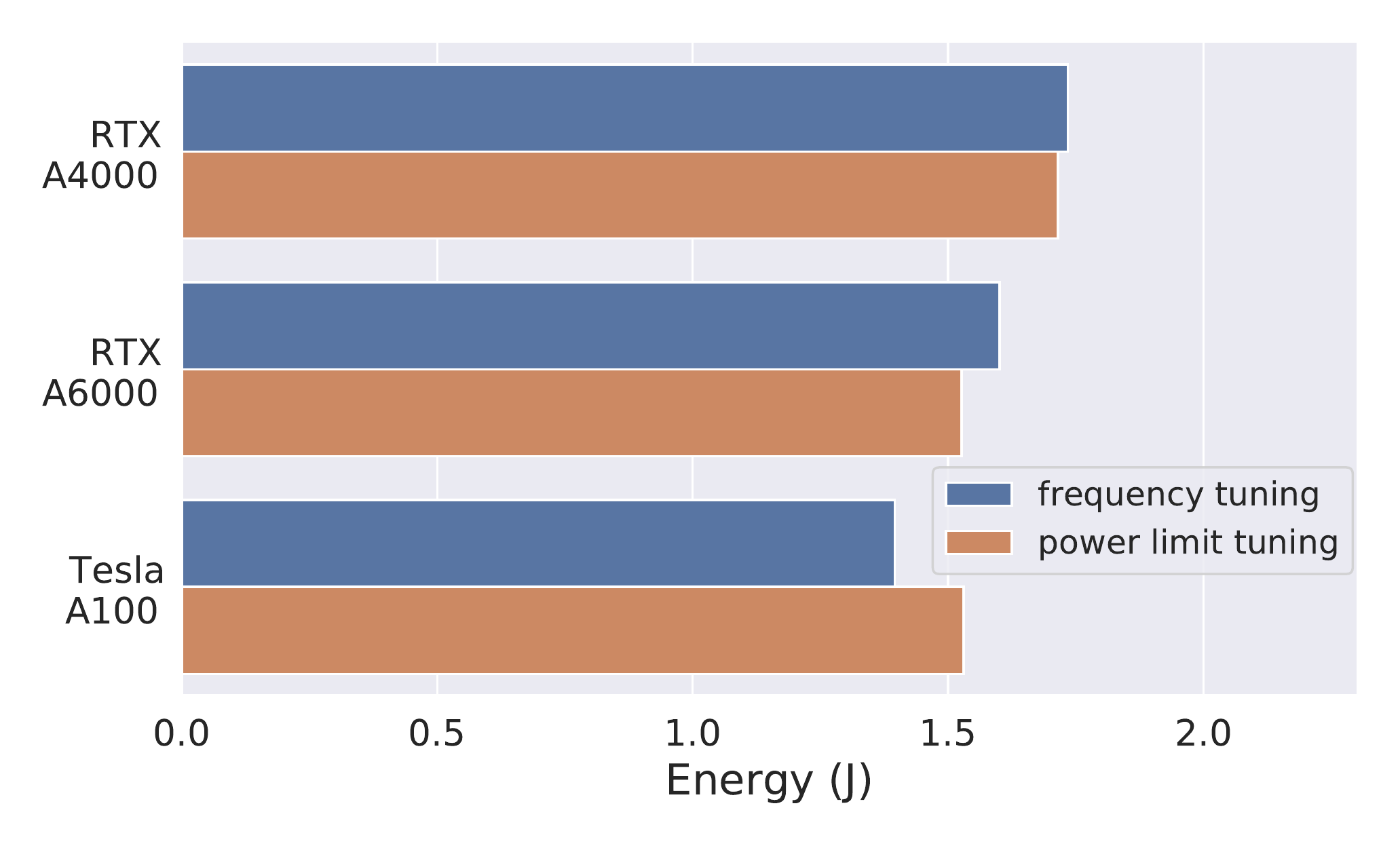}\\
    \vspace{-2.1em}
    \includegraphics[width=0.42\textwidth, center]{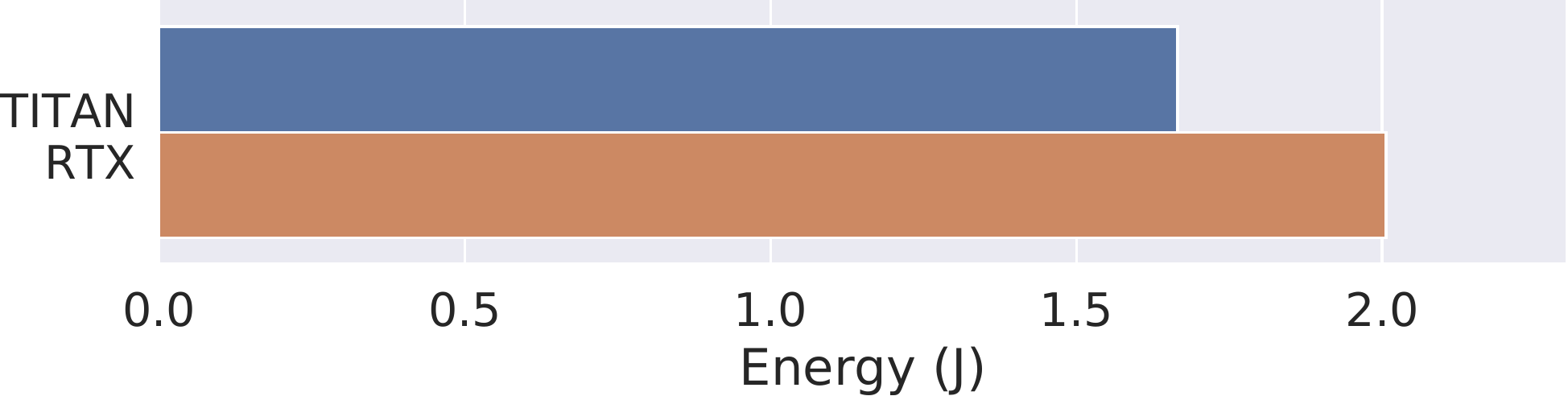}
    \vspace{1.0em}
    \caption{Lowest found energy for power capping or frequency tuning for GEMM, for the RTX A4000, RTX A6000, Tesla A100, and TITAN RTX GPUs. The energy measurements for the TITAN RTX were acquired using the PowerSensor2 instead of the NVML energy.}
    \label{fig:freq_vs_pwrlimit}
    \vspace{-0.6em}
\end{figure}

In Figure \ref{fig:FFG_GEMM}, we plot the proportion of centrality as a function of $p$ for GEMM, for the RTX A4000, RTX A6000, and Tesla A100 GPUs. For every GPU we plot the proportion of centrality curve for performance (time) tuning, energy tuning with clock frequency, and energy tuning with power limits. There does not appear to be a significant difference in difficulty for the RTX A4000 GPU. For the RTX A6000 GPU, the minima with more than $125\%$ runtime of the optimum are less central. However, as these minima are already significantly worse than the near-optimal solutions, we conclude that performance tuning is not significantly harder than energy tuning for the RTX A6000. For the Tesla A100, we find that energy tuning is significantly harder than performance tuning. For minima $\leq 110\%$ of optimal fitness, a local search algorithm is 2-4$\times$ less likely to terminate in these minima when minimizing energy.

Overall, in our experiments, energy tuning is either similar in tuning difficulty or harder depending on the GPU. As such, these search spaces remain infeasibly large to traverse fully within a day, and picking many sampling clock frequencies or power limits will compound this problem.

\begin{figure*}
    \centering
    \includegraphics[width=0.85\textwidth]{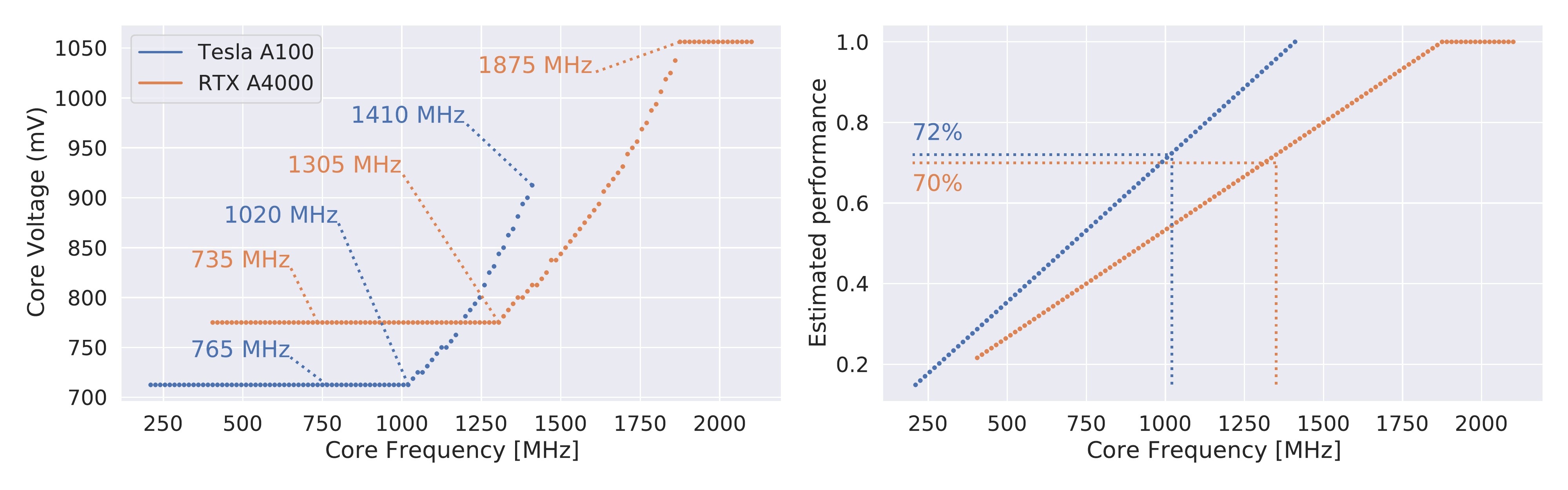}
    \caption{Left: GPU core frequency versus voltage curves for Tesla A100 and RTX A4000. The base clock frequency, the ridge point and peak frequency for each GPU are highlighted with a dashed line and label. Right: estimated performance under the assumption that GPU performance scales linearly with the clock frequency up to the point where throttling (if any) occurs. Estimated performance is normalized according to the performance for the highest possible clock frequency.}
    \label{fig:frequency_tuning_1}
    \vspace{-0.6em}
\end{figure*}

\subsection{Power capping versus frequency tuning}

In this section, we compare two methods that frequently appear in the literature; power capping~\cite{krzywaniak_performanceenergy_2019}, which is fixing the power limit of the GPU, and frequency tuning~\cite{mei_measurement_2013, ge_effects_2013, price_optimizing_2016, akiki_energy-aware_2018, fan_accurate_2020, calore_energy-performance_2015}, which aims to find the optimal application-specific GPU clock frequency.

In Figure~\ref{fig:power-vs-frequency-tuning}, we analyse
the impact of both frequency tuning and power capping on GPU power consumption. At the same measured frequencies, power consumption seems a bit higher when using a fixed clock frequency compared to setting a power limit. We observe that power capping does not cover the entire range of clock frequencies supported by the GPU. Therefore, using frequency tuning, we can reduce the power consumption below the minimum power limit, which may be beneficial for some applications. Moreover, by operating at a fixed clock frequency (below the point where throttling may occur), GPU behaviour is more predictable.

To compare the two methods globally, we add to the existing tunable GEMM parameters either a set of power limits or clock frequencies. We take a 7-point equidistant sample from the range of power limits in case of power capping, and the range of supported SM clock frequencies in case of frequency tuning. Using these parameters, we have performed a full combined search space exploration of the GEMM application on the RTX A4000, RTX A6000, Tesla A100 and TITAN RTX GPUs. On the Titan RTX, we measured power consumption using PowerSensor2 instead of NVML.

The lowest measured energy for power capping and frequency tuning is given in
Figure~\ref{fig:freq_vs_pwrlimit}. For the RTX A4000 and A6000 GPUs, power capping results in a marginally lower energy configuration, but not for the Tesla A100. For the TITAN RTX, where we used 20 sampling points for frequency tuning (300 MHz to 2100 MHz in steps of 75 MHz) and 9 for power capping (100 W to 300 W in steps of 25 W), we see that frequency tuning finds a significantly more energy efficient configuration. This seems to suggest that given sufficient sampling points, due to the increased frequency range, frequency tuning can result in a more energy efficient configuration. However, this leads to an increase in search points in an already large search space. To combat this, in Section~\ref{sec:powermodelling}, we investigate the relationship between frequency and voltage, and how this can be used to steer fine-grained frequency tuning. 

%The possible power limits for the A100, A4000, and A6000 are $150-250$, $100-140$, and $100-300$ respectively. However, the range of clock frequencies for the A100, A4000, and A6000 are $330-1410$, $570-2100$, and $390-2100$ respectively. The TITAN RTX results suggest that, with a larger range of frequencies, frequency tuning can result in a more energy efficient configuration. However, this leads to an increase in search points in an already large search space, which we will investigate in section \ref{sec:searchspace}. To combat this, in section \ref{sec:powermodelling}, we investigate the relationship between frequency and voltage, and how this can used to steer fine-grained frequency tuning. 

% In the next section, we look into modelling estimated power consumption to decrease the range of clock frequencies to be tested to find the most energy efficient solution.

\subsection{Model-steered frequency tuning}
\label{sec:powermodelling}
%The power consumption of a GPU is affected by several factors, including the workload and the operating frequency of the GPU. The workload is implementation dependent, and in most cases can be optimized by tuning kernel parameters or by changing the kernel code. The operating frequency, on the other hand, is more commonly taken as is.

%Contemporary GPUs usually operate at a base GPU core frequency and can boost up to a certain turbo frequency to increase performance, but only when the temperature and power consumption of the device allows for it. This technique is commonly referred to as Dynamic Voltage Frequency scaling (DVFS). Price et al. showed that there is a nonlinear relation between clock frequency and the voltage required to operate on a given frequency~\cite{price_optimizing_2016}. Consequently, the turbo frequency may be good for performance, but not necessarily for energy efficiency.

In this section, we analyse the impact of clock frequency scaling on the power consumption of the GPU, with the goal of identifying a range of suitable clock frequencies that likely results in energy-efficient configurations. The GPU core voltage can be queried by calling \texttt{NVIDIA-smi -q -d VOLTAGE}. In our experience, this option is only available with fairly recent NVIDIA drivers (510 and newer) in combination with Ampere GPUs (e.g. A100, A4000, A6000).

We plot the frequency-voltage curves for Tesla A100 and RTX A4000 in Figure~\ref{fig:frequency_tuning_1}. We observe that there is indeed a non-linear relation between core frequency and voltage, as discussed in Section~\ref{sec:powermodel}. For both the Tesla A100 and RTX A4000, the voltage remains unchanged for a range of core frequencies, after which the voltage increases seemingly quadratically. We will refer to the point where this increase occurs as the \emph{ridge point}.
The RTX A4000 seems to be capped at 1875 MHz, as the core voltage does not increase beyond this point. This is likely due to its power limit of 140W.
This is not observed for the Tesla A100, potentially due to its lower maximum operating frequency and higher power limit of 250W.
At the ridge points, the clock frequency for the GPUs is 72\% and 70\% of the peak clock frequency, for the Tesla A100 and RTX A4000 respectively. Interestingly, for both GPUs, the ridge point does not coincide with the base frequency.

\begin{figure*}
    \centering
    \includegraphics[width=0.9\textwidth]{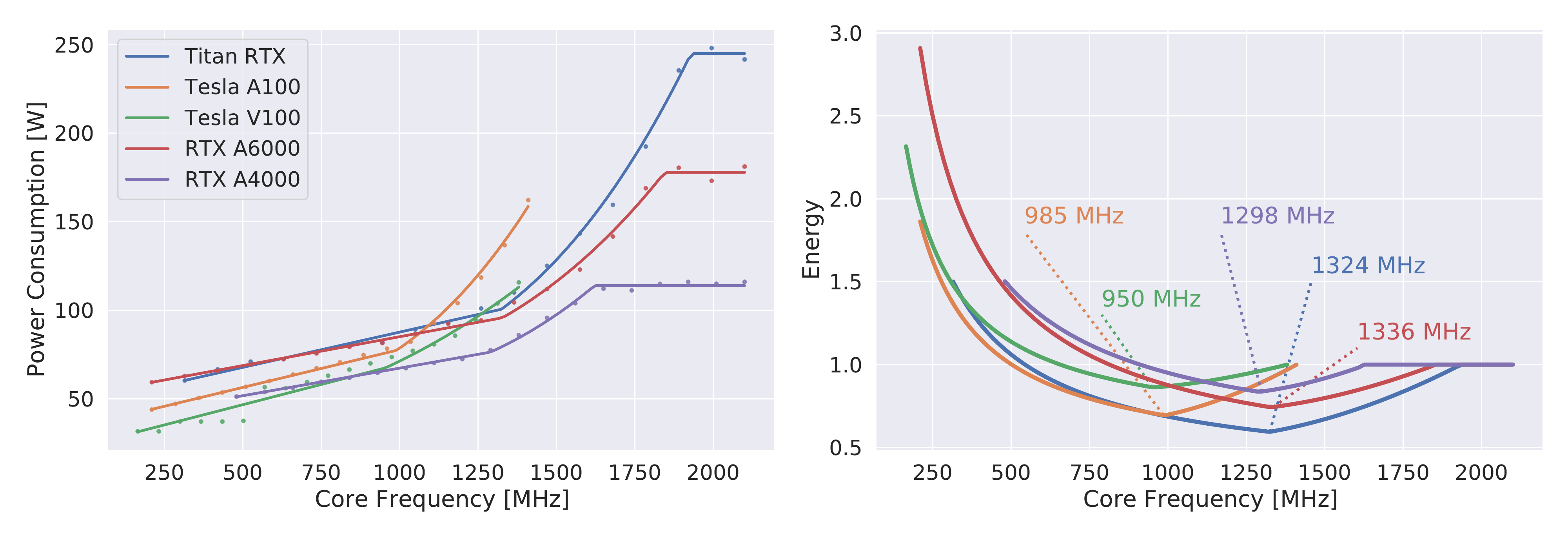}
    \vspace{-0.3em}
    \caption{Left: Power consumption of dot product kernel that fully loads the GPU, for the Tesla A100, RTX A4000, RTX A6000, Tesla V100, and Titan RTX. The dots indicate measurements, while the lines show the modelled power consumption (equation \ref{eq:powercurve}). Right: Corresponding estimated energy usage, with frequency that leads to minimal energy usage.}
    \label{fig:frequency_tuning_2}
    \vspace{-0.2em}
\end{figure*}

%While both the Tesla A100 and RTX A4000 are based on the Ampere architecture, we cannot assume that their energy-efficiency characteristics are the same. They use a different chip (GA100 versus GA102), are produced at a different process size (7 nm versus 8 nm), and have a very different mix and number of execution units. Moreover, the Tesla A100 has HBM2e memory, while the RTX A4000 uses GDDR6. The NVIDIA drivers currently do not expose an option to tune the clock frequency of the HBM memory. For A4000 and a compute-bound kernel, we measured only a marginally lower energy consumption when reducing the memory clock frequency. Therefore, our model is solely based on the graphics clock (core) frequency. By looking at two very distinct GPUs, we aim to provide a generic method for estimating the most efficient clock frequency that works for any contemporary GPU.

\subsubsection{Estimating GPU power consumption}
Equation \ref{eq:powersum} shows that the power consumption of a GPU can be modelled as the sum of the idle power and the dynamic power. In our model we take the idle power consumption as a constant, and the dynamic power consumption has a linear dependence on frequency, and a quadratic dependence on voltage. Moreover, for GPUs that are prone to power-limit throttling (e.g. RTX A4000), the power consumption of the GPU is capped. The model for estimated GPU power consumption is
\begin{equation}\label{eq:powercurve}
P_{load}^* = min(P_{max}, P_{idle}^* + \alpha * f * v^2).
\end{equation}
$P_{load}^*$, $P_{max}$, and $P_{idle}^*$ denote the estimated, maximum and idle power consumption of a GPU respectively. An initial value for $P_{max}$
can be obtained by measuring the maximum power consumption observed when executing  a kernel that fully loads the GPU, or simply by looking up the TDP of the device. $P_{idle}$ can be obtained by measuring the power consumption when no kernel is being executed. $\alpha$ is a constant, $f$ is the core frequency of the GPU, and $v$ denotes the GPU core voltage.

%\subsubsection{Estimating GPU performance}
% With a number of different GPU kernels (each with a fixed set of kernel parameters), we found that the runtime of the kernel scales inversely proportional to the core frequency.
%We estimate performance as a function of core frequency, taking into account that the core frequency may be capped due to the power limit.
%With (estimated) $P_{load}^*$ and $C_{f}$ as estimated performance, we compute estimated efficiency \cite{price_optimizing_2016} $\eta = C_{f} / P_{load}^*$.

\subsubsection{Estimating GPU core voltage}
For GPUs that do not support voltage readings, such as the Tesla V100 and Titan RTX, we extend the methodology outlined above to include a voltage estimate as a function of core frequency. We assume based on our observations that for these GPUs there exists a threshold $\tau_{ft}$ after which the voltage increases with a rate $\beta$. As input, our method requires a number of power measurements for a uniform sample of all the clock frequencies that the GPU supports. These data points are used to fit equation \ref{eq:powercurve} to estimate $P_{load}$, where $v$ is substituted by:
\begin{equation}
v(f) = \begin{cases}
  1  & f < \tau_{ft} \\
  \beta * (f-\tau_{ft}) & f >= \tau_{ft}
\end{cases}
\end{equation}

\subsubsection{Fitting the model}
We test our model by configuring Kernel Tuner to record core frequency and power usage while running a simple synthetic kernel (array dot product) that fully loads the GPU. We only need a few samples, spaced uniformly along the supported core frequencies. Using the measurements obtained with Kernel Tuner, for every GPU, we fit equation \ref{eq:powercurve} to the data as outlined in section \ref{sec:powermodel}. When fitting the model for $P^*_{load}$, the frequency $f$ runs till the highest clock frequency before throttling (if any) occurs.

\begin{table*}[]
    \centering
    \begin{tabular}{l|l|r r r r r r |r}
    \toprule
          & & \textbf{GOPs/W} & \textbf{GOPs/W} & \textbf{GOPs/W} & \textbf{TOP/s} & \textbf{TOP/s} & \textbf{TOP/s} & \textbf{Tuned} \\
        \textbf{GPU} & \textbf{Kernel} & \textbf{(before)} & \textbf{(after)} &  \textbf{gained} & \textbf{(before)} & \textbf{(after)} & \textbf{gained} & \textbf{frequency} \\
        \midrule
        \multirow{6}{*}{\emph{Tesla A100}}
 &                Gridder &     64.7 &    102.6 &        58.6\% &   16.3 &    12.0 &     -26.5\% & 1035 MHz\\
 &              Degridder &     59.8 &     97.5 &        63.1\% &   14.5 &    10.7 &     -26.2\% & 1035 MHz\\
 &        FD Dedispersion &     62.2 &     92.8 &        49.1\% &    9.7 &     7.3 &     -24.6\% & 1035 MHz\\
 &        TD Dedispersion &     13.3 &     21.5 &        61.3\% &    3.4 &     2.5 &    -26.4 \% & 1035 MHz\\
 & Tensor-Core Correlator &    684.8 &   1264.2 &        84.6\% &  148.4 &   135.2 &      -8.9\% & 1035 MHz\\
 &       LOFAR Correlator &     58.9 &    125.8 &       113.8\% &   12.2 &    10.7 &     -12.0\% & 1035 MHz\\
        \midrule
        \multirow{6}{*}{\emph{RTX A4000}}
 &                Gridder &     77.6 &    107.5 &        38.6\% &    11.0 &    8.1 &     -25.8\% & 1200 MHz\\
 &              Degridder &     90.8 &    131.6 &        44.9\% &    10.2 &    9.4 &      -8.1\% & 1470 MHz\\
 &        FD Dedispersion &     77.6 &    111.9 &        44.3\% &     8.3 &    6.7 &     -19.2\% & 1290 MHz\\
 &        TD Dedispersion &     12.9 &     17.2 &        33.0\% &     1.5 &    1.1 &     -22.2\% & 1200 MHz\\
 & Tensor-Core Correlator &    571.2 &    606.8 &         6.2\% &    57.2 &   55.2 &      -3.6\% & 1290 MHz\\
 &       LOFAR Correlator &     98.9 &    119.3 &        20.6\% &     8.7 &    8.4 &      -4.2\% & 1470 MHz\\
        \midrule
        \multirow{6}{*}{\emph{TITAN RTX}}
 &                Gridder &     55.2 &     68.6 &        24.2\% &    14.3 &    9.0 &     -37.2\% & 1260 MHz\\
 &              Degridder &     48.4 &     65.6 &        35.4\% &    13.7 &    8.2 &     -39.7\% & 1155 MHz\\
 &        FD Dedispersion &     39.9 &     59.9 &        50.2\% &    10.2 &    5.5 &     -45.4\% & 1050 MHz\\
 &        TD Dedispersion &      8.0 &     12.1 &        50.7\% &     2.1 &    1.3 &     -40.0\% & 1050 MHz\\
 & Tensor-Core Correlator &    140.5 &    209.5 &        49.1\% &    34.7 &   23.4 &     -32.6\% & 1155 MHz\\
 &       LOFAR Correlator &     51.5 &     78.0 &        51.6\% &    12.8 &    7.2 &     -43.4\% & 1155 MHz\\
        \midrule
        \multirow{6}{*}{\emph{Tesla V100}}
 &                Gridder &     59.6 &     73.6 &        23.6\% &    11.6 &    9.5 &     -18.0\% & 1110 MHz\\
 &              Degridder &     61.7 &     74.2 &        20.2\% &    11.0 &    8.8 &     -19.9\% & 1110 MHz\\
 &        FD Dedispersion &     58.6 &     69.2 &        18.1\% &     7.4 &    6.0 &     -19.2\% & 1110 MHz\\
 &        TD Dedispersion &     11.6 &     15.7 &        34.9\% &     2.2 &    1.3 &     -37.8\% & 1110 MHz\\
 & Tensor-Core Correlator &    260.8 &    301.5 &        15.6\% &    34.2 &   27.7 &     -18.9\% & 1110 MHz\\
 &       LOFAR Correlator &     74.7 &     86.8 &        16.3\% &     9.9 &    7.6 &     -23.5\% & 1110 MHz\\
   \bottomrule
    \end{tabular}
    \caption{Energy efficiency (GOPs/W) and compute performance (TOP/s) before and after model-steered frequency tuning, i.e., select the most energy-efficient frequency within $\pm 10\%$ MHz of the ridge points found in Figure \ref{fig:frequency_tuning_2}. All kernels use floating point operations (FLOPs) except the Tensor-Core correlator, which uses 16-bit integer operations. \\
    $^*$\textbf{Note:} The before measurements are already tuned for time by a domain expert.}
    \label{tab:LOFARenergy}
    \vspace{-0.6em}
\end{table*}

The left plot in Figure~\ref{fig:frequency_tuning_2} illustrates that the estimated power consumption closely follows the power consumption measured using NVML. Next, the estimated power consumption is used to compute estimated energy usage as a function of absolute power ($P_{load}^*$) divided by clock frequency ($f$). For each of the GPUs, there is a core frequency that minimizes estimated energy usage, see Figure~\ref{fig:frequency_tuning_2} (right). For both the Tesla A100 and RTX A4000, the predicted most energy-efficient clock frequencies (985 MHz and 1298 MHz) are close to the observed ridge points at 1025 MHz and 1290 MHz as identified in Figure~\ref{fig:frequency_tuning_1}.

Reducing the clock frequency beyond the ridge point does not make the GPU more energy efficient, as performance drops with $f$ while $v$ is constant below the ridge point. This leads to a higher total energy usage for non-zero $P_{idle}$. On the other hand, there is a trade-off between performance and energy  when considering higher clock frequencies than the ridge point, up to the point where throttling starts to occur (at about 1700 MHz for the RTX A4000 and 2000 MHz for Titan RTX).
As energy increases quadratically with voltage, and compute performance linearly with frequency, it is unnecessary to consider frequencies significantly higher than the ridge point.
%Therefore, for the A4000, we would recommend running Kernel Tuner for frequencies in the range of 1200 MHz up to 1700 MHz. This is only a third of the full range of frequencies that this GPU supports, and will therefore considerably reduce the tuning time.

\begin{figure*}
    \centering
    \includegraphics[width=\textwidth]{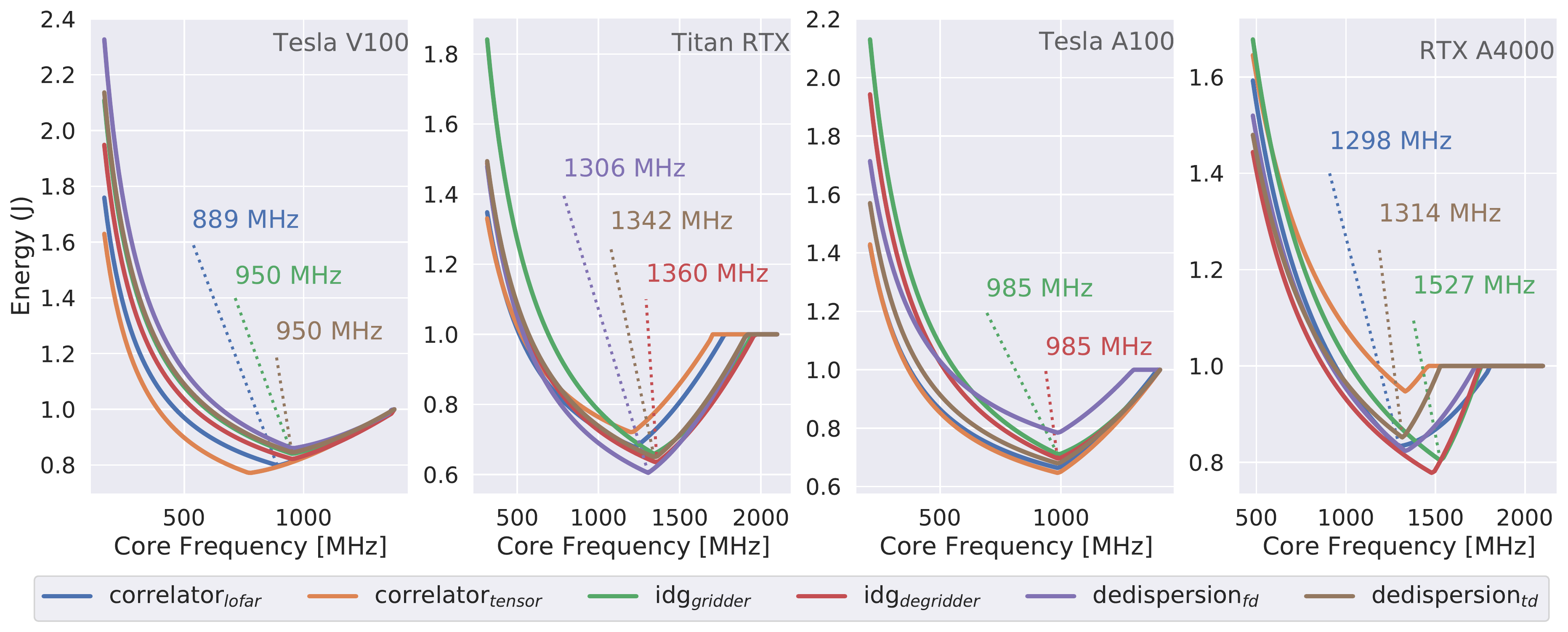}
    \caption{Modelled energy usage (J) with power consumption model for core clock frequencies (MHz) of LOFAR kernels for the Tesla V100, Titan RTX, Tesla A100 and RTX A4000 GPUs.}
    \label{fig:frequency_tuning_4all}
    \vspace{-0.6em}
\end{figure*}

%To verify that the method works in general, we have conducted the same analysis for a range of different kernels shown in Figures \ref{fig:frequency_tuning_3a} and \ref{fig:frequency_tuning_3b} for A100 and A4000 respectively. In both cases, our model fits estimated power consumption close to the measured power consumption. For A100, the optimum frequency seems to be 985 MHz for all kernels. For A4000, there is a range of \emph{best} core frequencies: ~1250 Mhz up to ~1450 Mhz. This lower bound coincides with the estimate of 1271 MHz for the ridge point.

%Finally, we compare the results of our method to an exhaustive search in Figure~\ref{fig:frequency_tuning_4}. For A100, 1025 MHz (estimated ridge point at 985 MHz) is the most energy efficient frequency for all applications, and for A4000 this is 1300 MHz (estimated ridge point at 1271 MHz).
%
%The most energy-efficient frequencies are within $5\%$ of the predicted frequencies.
%
%This shows that our power consumption model is capable of determining the approximate most energy-efficient clock frequency.

To conclude, prior to energy tuning a particular GPU kernel, we recommend running a kernel that fully loads the GPU for a range of clock frequencies. Our model can then be used to fit a power consumption curve and find an estimate for the most energy-efficient frequency. Next, energy tuning can be run with a fine-grained sampling of clock frequencies around the estimated optimal frequency. This feature is included in Kernel Tuner\footnote{https://github.com/KernelTuner/kernel\_tuner} (version 0.4.4). In this work, we use a range of $\pm 10\%$ of the optimal frequency estimated with the model.
%According to our validation, this method provides a fast and simple way to save up to 40\% energy while only costing about 25\% performance.

\iffalse
\begin{figure*}
    \centering
    \includegraphics[width=\textwidth]{frequency-tuning-3a.pdf}
    \caption{Left: Measured power consumption on the Tesla A100 GPU (dots) for clock frequencies of LOFAR kernels, with modelled power consumption (line). These kernels are previously tuned for compute performance. Right: Corresponding estimated energy usage. The same clock frequency of 985 MHz minimizes energy to solution.}
    \label{fig:frequency_tuning_3a}
    \vspace{-0.5cm}
\end{figure*}

\begin{figure*}
    \centering
    \includegraphics[width=\textwidth]{frequency-tuning-3b.pdf}
    \caption{Left: Measured power consumption on RTX A4000 GPU (dots) for clock frequencies of LOFAR kernels, with modelled power consumption (line). These kernels are previously tuned for compute performance. Right: Corresponding estimated energy usage. The most energy-efficient clock frequency differs per application.}
    \label{fig:frequency_tuning_3b}
    \vspace{-0.5cm}
\end{figure*}

\begin{figure*}
    \centering
    \includegraphics[width=\textwidth]{frequency-tuning-4.pdf}
    \caption{Measured energy usage for clock frequencies for exhaustive search of dot product kernel that fully loads the GPU, for Tesla A100 (left) and RTX A4000 (right) GPUs. The most energy-efficient frequency found is designated.}
    \label{fig:frequency_tuning_4}
    \vspace{-0.5cm}
\end{figure*}
\fi

\subsection{Practical efficiency gain for radio astronomy kernels}
\label{sec:practicalgain}
To verify the energy gains on a real-world high-throughput pipeline, we apply our model-steered frequency tuning method to the six radio astronomy LOFAR kernels (see section \ref{sec:setup}) currently running on the DAS-6 system \cite{bal2016das}, and LOFAR COBALT-2 system~\cite{broekema2018cobaltcorrelator} (can receive more than 1 Tbit/s). By using model-steered frequency tuning we reduce the size of the searchspaces by 82.4\%, 78.9\%, 77.8\%, and 80.0\% for the Tesla A100, RTX A4000, Titan RTX, and Tesla V100 respectively. The measured compute performance and energy efficiency before and after model-steered tuning is given in Table \ref{tab:LOFARenergy}. Note that all six kernels have previously been optimized for compute performance, which means that the reduction in compute performance may be more severe than in most cases.

After model-steered frequency tuning, the LOFAR kernels gained between $\sim$15--113\% in energy efficiency, while losing $\sim$3--45\% compute performance. Gains in energy efficiency, and losses in compute performance, varied significantly between GPU models and kernels.
Two notable outliers are the Tensor-Core correlator on the RTX A4000, where efficiency increased only 6\%, and the LOFAR correlator on the Tesla A100, where an efficiency gain of 113.8\% was achieved while losing only 12\% compute performance. Overall, the mean energy efficiency gain was $42.0\pm 24.1$\%, and the mean compute performance loss was $-24.3\pm 12.1$\%.
%Overall, these results show that significant gains in energy efficiency are possible using model-steered frequency tuning, while avoiding a full brute force search over the energy tuning space.

The estimated energy usage curves for each application using the power consumption model are given in Figure~\ref{fig:frequency_tuning_4all}. We can see that sometimes the estimated optimal frequency is close to the measured optimal frequency in Table \ref{tab:LOFARenergy}, and sometimes differs more significantly. In future work, we plan to expand the model by adding memory- and temperature-dependent terms.
%While losing compute performance may be detrimental for certain applications,
%Note that many large-scale computing systems are significantly larger than necessary and have hardware redundancies. This is also the case for the COBALT-2 system (25\% extra GPUs). Therefore, the system may suffer a less severe loss in compute performance than measured here on individual GPUs.
%This suggests that significant energy efficiency gains can potentially be made on such systems with small reductions in compute performance.
%However, as the COBALT-2 system is complex and contains bottlenecks, a system-level analysis would have to be performed to accurately determine these metrics.

\section{Conclusions}

We have investigated several GPU kernel tuning approaches for improving energy efficiency, and extended Kernel Tuner's capabilities for measuring GPU power consumption and for tuning energy usage. %Using these methods, we have compared several different optimization strategies that are commonly applied in the literature.
On a commonly-used benchmark matrix multiplication kernel (GEMM) -- designed for compute performance without energy-specific tunable parameters -- we found that with a speed reduction of $27.5\%$ an increase in energy efficiency of $50.9\%$ is possible on the Tesla A100.
Additionally, the combined search space of all tunable parameters (including clock frequency) contains a globally lower energy configuration, compared to tuning for performance and then tuning clock frequency separately. However, for most GPUs tuning the frequency separately did lead to a close to optimal energy usage. When investigating energy tuning methods, we found that clock frequency tuning gives more fine-grained control over GPU power consumption than power capping, and enables a larger (and lower) range of power consumption.
%
%In addition, we analyzed the tuning difficulty of energy tuning as an optimization problem and found that it is similar to, or harder than, tuning for kernel runtime as an optimization problem.

Due to the prohibitively large search spaces when tuning both kernel parameters and clock frequency, we introduced a model to estimate GPU power consumption. We show that a single core clock frequency is the most energy efficient when the other tunable parameters are constant. This clock frequency can easily be estimated using our power consumption model. We verified the potential energy efficiency gains when tuning around $\pm10\%$ of our estimated frequency on a number of real-world radio astronomy kernels, and increased energy efficiency more than two fold at a loss of 12\% compute performance. Overall, the mean energy efficiency gain was $42.0\pm 24.1$\%, and the mean compute performance loss was $-24.3\pm 12.1$\%.
Using our model-steered frequency tuning approach, we were able to dramatically reduce the size of the auto-tuning search spaces by $77.8 - 82.4\%$.
%Using our model-steered frequency tuning approach, the energy efficiency of GPU systems can be significantly increased without requiring large tuning time.

\section*{Acknowledgment}
This research was carried out within the CORTEX project, funded by the Dutch Research Council (NWO) in the framework of the NWA-ORC Call (file number NWA.1160.18.316).
The DAS-6 cluster is funded through NWO-M and Open Competition (617.001.204) grants.

%\clearpage

\bibliographystyle{IEEEtran}
\bibliography{main_rewrite}

\end{document}

%% file: code-listings.tex
\usepackage{listings}

\newcommand\cudastyle{\lstset{ %
language=C,                   % choose the language of the code
numbers=left,                 % where to put the line-numbers
numberstyle=\scriptsize,      % the size of the fonts that are used for the lin$
stepnumber=1,                 % the step between two line-numbers. If it is 1$
numbersep=5pt,                % how far the line-numbers are from the code
showspaces=false,             % show spaces adding particular underscores
showstringspaces=false,       % underline spaces within strings
showtabs=false,               % show tabs within strings adding particular un$
frame=none,                   % frame around the code
otherkeywords={__global__,__shared__,__constant__}, % Add keywords 
xleftmargin=10pt,       %
breakatwhitespace=false,  %
tabsize=2,                    % sets default tabsize to 2 spaces
captionpos=b,                 % sets the caption-position to bottom
breaklines=true,              % sets automatic line breaking
breakatwhitespace=false,      % sets if automatic breaks should only happen at wh$
escapeinside={\#}{\#},        % if you want to add LaTeX within your code
linewidth=\linewidth,     %
basicstyle=\scriptsize\ttfamily, %
}}

\lstnewenvironment{cuda}[1][]
{
\cudastyle
\lstset{#1}
}
{}

\lstnewenvironment{opencl}[1][]
{
\cudastyle
\lstset{#1}
}
{}

\newcommand\pythonstyle{\lstset{ %
language=Python,              % choose the language of the code
numbers=left,                 % where to put the line-numbers
numberstyle=\scriptsize,      % the size of the fonts that are used for the lin$
stepnumber=1,                 % the step between two line-numbers. If it is 1$
numbersep=5pt,                % how far the line-numbers are from the code
showspaces=false,             % show spaces adding particular underscores
showstringspaces=false,       % underline spaces within strings
showtabs=false,               % show tabs within strings adding particular un$
frame=none,                   % frame around the code
xleftmargin=10pt,       %
breaklines=true,
tabsize=4,                    % sets default tabsize to 4 spaces
captionpos=b,                 % sets the caption-position to bottom
linewidth=\linewidth,     %
basicstyle=\scriptsize\ttfamily, %
}}

\lstnewenvironment{python}[1][]
{
\pythonstyle
\lstset{#1}
}
{}